\documentclass[fleqn,usenatbib]{mnras}

\usepackage[T1]{fontenc}

\usepackage{graphicx}	
\usepackage{amsmath}	
\usepackage{amssymb}	

\usepackage{newtxtext,newtxmath}

\usepackage{booktabs}

\usepackage{bm} 

\usepackage{siunitx}
\DeclareSIUnit\Rjup{\ensuremath{\mathit{R}}_\mathrm{J}}
\DeclareSIUnit\Mjup{\ensuremath{\mathit{M}}_\mathrm{J}}

\DeclareSIUnit\Rsun{\ensuremath{\mathit{R}_{\sun}}}
\DeclareSIUnit\Msun{\ensuremath{\mathit{M}_{\sun}}}
\DeclareSIUnit\Lsun{\ensuremath{\mathit{L}_{\sun}}}
\DeclareSIUnit\rhosun{\ensuremath{\rho_{\sun}}}

\DeclareSIUnit\Mearth{\ensuremath{\mathit{M}_{\earth}}}
\DeclareSIUnit\Rearth{\ensuremath{\mathit{R}_{\earth}}}

\DeclareSIUnit\Rstar{\ensuremath{\mathit{R}_{\star}}}

\DeclareSIUnit\ppm{ppm}
\DeclareSIUnit\ppt{ppt}

\DeclareSIUnit\day{d}
\DeclareSIUnit\dday{day}
\DeclareSIUnit\days{days}

\DeclareSIUnit\cgs{cgs }
\DeclareSIUnit\logg{\ensuremath{\log{(\si{\centi\metre\per\square\second})}}}
\DeclareSIUnit\dex{dex}
\DeclareSIUnit\hjdutc{\ensuremath{\mathrm{HJD}_\mathrm{UTC}}}
\DeclareSIUnit\bjdutc{\ensuremath{\mathrm{BJD}_\mathrm{UTC}}}
\DeclareSIUnit\bjdtdb{\ensuremath{\mathrm{BJD}_\mathrm{TDB}}}
\DeclareSIUnit\au{AU}
\DeclareSIUnit\pc{pc}

\DeclareSIUnit\yr{year}
\DeclareSIUnit\yr{yr}
\DeclareSIUnit\Gyr{Gyr}
\DeclareSIUnit\Myr{Myr}

\newcommand{\textblue}[1]{\textcolor{blue}{#1}}

\newcommand\ccfin{CCF$_\mathrm{in}$}
\newcommand\ccfout{CCF$_\mathrm{out}$}

\newcommand\istar{\ensuremath{i_\star}}

\newcommand*{\ra}[2][]{{
    \def\SIUnitSymbolDegree{\textsuperscript{h}}%
    \def\SIUnitSymbolArcminute{\textsuperscript{m}}%
    \def\SIUnitSymbolArcsecond{\textsuperscript{s}}%
    \ang[#1]{#2}}%
}

\newcommand\dnu{\ensuremath{\Delta\nu}}
\newcommand\vmax{\ensuremath{\nu_\mathrm{max}}}

\newcommand\pimen{\ensuremath{\pi} Men}
\newcommand\pimenb{\ensuremath{\pi} Men\,b}
\newcommand\pimenc{\ensuremath{\pi} Men\,c}
\newcommand\espresso{\emph{ESPRESSO}}
\newcommand\tess{\emph{TESS}}
\newcommand\ellc{\textsf{ellc}}
\newcommand\emcee{\textsf{emcee}}
\newcommand\celerite{\textsf{celerite}}
\newcommand\exoplanet{\textsf{exoplanet}}
\newcommand\pymcthree{\textsf{pymc3}}
\newcommand\vsini{\ensuremath{v_\mathrm{eq}\sin{i_\star}}}

\title[Orbital misalignment of the super-Earth \ensuremath{\pi} Men\,c]{
Orbital misalignment of the super-Earth \ensuremath{\bm{\pi}} Men\,c with the spin of its star}

\author[V. Kunovac Hod\v{z}i\'c et al.]{%
Vedad Kunovac Hod\v{z}i\'c$^{1}$%
\thanks{Contact: \href{mailto:vxh710@bham.ac.uk}{vxh710@bham.ac.uk}},
Amaury H.\,M.\,J. Triaud$^{1}$, 
Heather M. Cegla$^{2,3}$\thanks{CHEOPS Fellow, SNSF NCCR-PlanetS},
William J. Chaplin$^{1,4}$\newauthor
and Guy R. Davies$^{1,4}$
\\
\\
$^{1}$School of Physics and Astronomy, University of Birmingham, Edgbaston, Birmingham B15 2TT, UK \\
$^{2}$Department of Physics, University of Warwick, Gibbet Hill Road, Coventry CV4 7AL, United Kingdom\\
$^{3}$Observatoire de Gen\`eve, Universit\'e de Gen\`eve, Chemin des Maillettes 51,
1290 Sauverny, Switzerland \\
$^{4}$Stellar Astrophysics Centre (SAC), Department of Physics and Astronomy, Aarhus University, Ny Munkegade 120, DK-8000 Aarhus C, Denmark\\
}

\date{Accepted XXX. Received YYY; in original form ZZZ}

\pubyear{2020}

\begin{document}
\label{firstpage}
\pagerange{\pageref{firstpage}--\pageref{lastpage}}
\maketitle

\sisetup{separate-uncertainty=true, multi-part-units=single}

\begin{abstract}
    Planet-planet scattering events can
    leave an observable trace of a planet's migration history in the form of orbital
    misalignment with respect to the stellar spin axis, which is measurable from spectroscopic timeseries taken during transit. 
    We present 
    high-resolution spectroscopic transits observed with \espresso{} of the close-in super-Earth  \pimenc{}. The system also contains an outer giant planet on a
    wide, eccentric orbit, recently found to be inclined with respect to the inner planetary orbit. These characteristics are reminiscent of past dynamical interactions. We successfully retrieve the planet-occulted light during transit,
    and find evidence that the orbit of \pimenc{} is moderately misaligned with the stellar spin axis with $\lambda = \ang[angle-symbol-over-decimal]{-24.0} \pm \ang[angle-symbol-over-decimal]{4.1}$ 
    ($\psi = \ang[angle-symbol-over-decimal]{26.9}^{+5.8^\circ}_{-4.7^\circ}$). 
    This is consistent with the super-Earth \pimenc{} having followed a high-eccentricity migration followed by tidal circularisation, and hints that super-Earths can form at large distances from their star.
    We also detect clear signatures of solar-like oscillations within our \espresso{} radial velocity timeseries, where we reach a radial velocity precision of \SI{\sim20}{\centi\metre\per\second}. We model the oscillations using Gaussian processes and retrieve a frequency of maximum oscillation, $\vmax{} = 2771^{+65}_{-60}\,\si{\micro\hertz}$. These oscillations makes it challenging to detect the Rossiter-McLaughlin effect using traditional methods. We are, however, successful using the {\it reloaded} Rossiter-McLaughlin approach. Finally, in an Appendix we also present physical parameters and ephemerides for \pimenc{} from a Gaussian process transit analysis of the full \tess{} Cycle 1 data. 
    
\end{abstract}



\section{Introduction}
Perhaps one of the most surprising results from two decades of exoplanet
research is that planet sizes between that of Earth and Neptune
(\SIrange{1.5}{2}{\Rearth}) are the most likely outcome of planet formation
(e.g. \citealt{borucki2010,batalha2013}), even when such planets are completely
absent from our own Solar System.  Dubbed \emph{super-Earths} (see
\citealt{schlichting2018} for a review), these planets orbit \SI{50}{\percent}
of Sun-like (FGK) stars \citep{howard2010,mayor2011,fressin2013}, and rises to
\SI{75}{\percent} when including M dwarfs 
\citep{bonfils2013,dressing2015,gaidos2016,hardegree-ullman2019}. The
currently operating \tess{} survey \citep{ricker2015} is expected to find
\num{\sim 1000} additional super-Earths and mini-Neptunes in its 2-year
nominal mission lifetime \citep{barclay2018,huang2018a}.

Yet for their abundance, there have been few observational constraints on their
formation and dynamical evolution. In general, super-Earth formation consists of
core formation followed by gas accretion onto the assembled core
\citep{pollack1996,chabrier2014}.  In the context of \emph{in-situ} formation,
the inner protoplanetary disk does not have enough solid material in close-in
feeding zones, such that any embryo will reach an isolation (maximum) mass well
below that of super-Earths \citep{armitage2013}.
Core accretion until isolation mass followed by a giant impact phase is also
unlikely to produce the observed super-Earth population \citep{hansen2012,
schlichting2014,dawson2016}, which points to another likely mechanism at work to
explain the close in super-Earths. Several super-Earths have been detected in circumbinary configurations \citep[e.g.][]{orosz2019,kostov2020}. Due to strong gravitational perturbations produced by the binary orbital motion onto protoplanetary discs, planet formation is thought to only be possible at several AU from the central binaries \citep[e.g.][]{paardekooper2012,pierens2020} implying that the detected systems had to migrate in \citep{martin2018,pierens2020}.


Close-in planet formation may be aided by an influx of solids in the form of
pebbles \citep{johansen2017,lambrechts2019}, planetesimals,  or even fully
formed cores from a mass reservoir at several AU, where the isolation mass is
higher. In either scenario, \emph{in-situ} formation or inwards gas disk
migration (see \citealt{baruteau2014} for a review), planets would remain
aligned with the stellar equator, even in the event of interactions in
multi-planet systems \citep{bitsch2013}. 



This is not the case for high-eccentricity migration, which can deliver fully
formed super-Earths (or even hot Jupiters) to close-in, \emph{inclined} orbits
from beyond the snowline. 
In this two-step process, an outer giant planet will scatter the super-Earth to
a highly elliptical orbit. At each periastron passage, the super-Earth will pass
very close to the star (a few hundreths of AU) and cause tidal dissipation in
the planet, 
which acts to reduce the orbital energy and circularise the orbit to its present
day location.  The initial scattering event produces a broad distribution of
orbital inclinations \citep{fabrycky2007,naoz2011a,chatterjee2008}, unlike the
aforementioned disk migration pathways that results in co-planar systems.

One of the most promising ways of distinguishing different migration pathways is
through the Rossiter-McLaughlin effect \citep{rossiter1924,mclaughlin1924},
which measures the projected spin-orbit angle between the orbital plane and
stellar spin axis, $\lambda$.  Rossiter-McLaughlin measurements have all but
become routine observables for hot Jupiters, and a trend has emerged in which
stars above the Kraft break ($T_\mathrm{eff} \gtrsim \SI{6200}{\kelvin}$) tend
to host misaligned massive planets, while stars below (cold stars) tend to have
co-planar planets (see \citealt{triaud2018} for a review, and references
therein). In general this can be explained by hot Jupiters being massive enough
to realign the stellar spin axis from tidal coupling to the their thick
convective envelopes, and spin-down due to magnetic breaking, while more massive
stars have thinner -- or non-existent -- convective envelopes and remain fast
rotators \citep{winn2010a,dawson2014}. These considerations muddle our
interpretation of the dynamical histories of exoplanets, and thus inferring the
migration pathways and formation mechanisms of close-in massive planets from
spin-orbit angle measurements therefore remains difficult. Smaller planets,
however -- such as super-Earths -- are not massive enough to realign the star
within the tidal decay timescale or lifetime of the star \citep{dawson2014}, and
will therefore keep their orbital inclinations, allowing us to robustly infer
their migration pathways from their present-day orbital obliquities.  However,
Rossiter-McLaughlin observations have thus far largely eluded the super-Earths.
The expected radial velocity semi-amplitudes are often at or below the m/s
level, which is at the precision limit of our best spectrographs on the
brightest stars, and where phenomena such as stellar oscillations and near-surface convection  begin to dominate the signal. Nevertheless,
successful Rossiter-McLaughlin campaigns have been carried out on small planets,
such as the misaligned Neptune-mass exoplanets HAT-P-11\,b \citep{winn2010c,
hirano2011}, GJ 436$\,$b \citep{bourrier2018}, the somewhat contentious
result on the super-Earth 55 Cnc$\,$e \citep{bourrier2014,lopez-morales2014}, and most recently on the 22 Myr Neptune-sized planet AU Mic\,b \citep{addison2020,hirano2020a,palle2020}.

Here, we present a detection of the planetary shadow of a close-in super-Earth,
known to host a giant planet companion on a wide, eccentric orbit. We find that
the orbit is misaligned with the stellar spin-axis, pointing to an origin beyond
the snowline and may show evidence of high-eccentricity migration following
dynamical interaction with the outer companion in the system's youth.  We
present spectroscopic transits obtained with the \espresso{} spectrograph
\citep{pepe2014}, as one of the first science observations after completing its
commissioning, and demonstrate its capabilities for providing observational
constraints on planet formation of the most common population of exoplanets.

The paper is organised as follows: 
In Section~\ref{section:espresso} we present our analyses of the \espresso{} data that includes \emph{i)} modelling of the
Rossiter-McLaughlin anomaly combined with a Gaussian process model for the
asteroseismic activity, and \emph{ii)} an independent analysis using the reloaded
Rossiter-McLaughlin method. In addition, in Appendix~\ref{appendix:tess} we include our transit analysis using the full \tess{} Cycle 1 data to inform our Rossiter-McLaughlin modelling. In Section~\ref{section:results} we present our updated transit parameters, and new asteroseismic parameters and obliquity measurements. We discuss our results in the context of planet formation in Section~\ref{section:discussion},
and conclude in Section~\ref{section:conclusion}. In Appendix~\ref{appendix:tess} we present our
analysis on \tess{} photometric data to constrain the transit ephemerides and
parameters for the Rossiter-McLaughlin analysis.

\section{\espresso{} data analysis}
\label{section:espresso}

\subsection{Observations and data reduction}

Two transits of \pimenc{} were observed on the nights of 2 November
2018 (run A) and 16 December 2018 (run B) using the \espresso{} spectrograph
\citep{pepe2014} 
mounted on the Very Large Telescope at ESO Paranal Observatory (DDT
2102.C-5008, PI: Hod\v{z}i\'c). The transit in run A was observed concurrently
with \tess{}  Sector 4 observations. \espresso{} observations were carried out under very good
conditions (seeing ${\sim}\SI{0.5}{\arcsec}$) in the high-resolution mode using
1x1 binning. 
Integration times were fixed at \SI{120}{\second} and \SI{80}{\second} for run A
and B, respectively,
with \SI{44}{\second} dead-time per observation due to read-out, reaching median SNR
per pixel of $272$ and $200$ at \SI{550}{\nano\metre} for each run. 
The two runs cover respectively \SI{5}{\hour} and \SI{6}{\hour} uninterrupted
sequences that cover the full transit duration and additional 
baselines of \SIrange{2}{3}{\hour} before and after the transits.
The observations are summarised in Table~\ref{tab:observations} (\textit{top}).

\begin{table}
\renewcommand{\arraystretch}{1.2}
    \centering
    \caption{Summary of \espresso{} and \tess{} data used in this work.}
    \resizebox{\columnwidth}{!}{
    \begin{tabular}{@{\extracolsep{\fill}}
            ccccccc}
        \multicolumn{7}{c}{\espresso{} (Section~\ref{section:espresso})}\\
        \toprule
        \toprule
        {ESO ID} & {Run} & {Night} & {$N_\mathrm{obs}$} & {$t_\mathrm{exp}$} & 
        {$\langle \mathrm{SNR} \rangle$} &
        {$\langle \sigma_\mathrm{RV} \rangle$}\\
        & & & & (s) & & (\si{\centi\metre\per\second}) \\
        \midrule 
        2012.C-5008(A) & A & 2018-11-02 & 111 & 120 & 272 &
        24  \\
        2012.C-5008(B) & B & 2018-12-16 & 171 & 80 & 200 &
        35  \\
        \bottomrule
        \vspace{2pt}
    \end{tabular}}
   \resizebox{\columnwidth}{!}{
        \begin{tabular*}{\columnwidth}{@{\extracolsep{\fill}}
        ccccc}
        \multicolumn{5}{c}{\tess{} (Appendix~\ref{appendix:tess})}\\
        \toprule
        \toprule
        {Sector} & {Date} & {$N_\mathrm{obs}$} & {$t_\mathrm{exp}$} & 
        {$\sigma_\mathrm{residual}$} \\
        & & & (s) & (ppm) \\
        \midrule 
        1 & 25 July -- 22 August 2018 & \num{18054} & 120 & 124 \\
        4 & 18 October -- 15 November 2018 & \num{15768} & 120 &  114 \\
        8 & 2 -- 28 February 2019 & \num{13395} & 120 & 133 \\
       11 & 22 April -- 21 May 2019 & \num{15299} & 120 &  120 \\
       12 & 21 May -- 19 June 2019 & \num{19071} & 120 &  137 \\
       13 & 19 June -- 18 July 2019 & \num{19562} & 120 &  110 \\
        \bottomrule
    \end{tabular*}}
    \label{tab:observations}
\end{table}

The spectra were reduced with version 2.0 of the \espresso{} data reduction
pipeline\footnote{\url{ftp://ftp.eso.org/pub/dfs/pipelines/espresso/espdr-reflex-tutorial-1.0.0.pdf}}, 
using a G2 binary mask to create cross-correlation functions (CCFs) that are
fitted with Gaussian profiles. From the Gaussian fits we derive the depth (contrast),
width (FWHM), and mean (radial velocity). The first sequence shows some
variation in the derived contrast and SNR throughout the night which may
be due to passing thin clouds. In Section~\ref{section:rmreloaded_description} we describe a way of mitigating its impact on our result.
The second sequence is stable with the contrast normally
dispersed around the mean, but with a slope as the SNR increases throughout the night. 
The resulting median uncertainties on the integrated radial
velocities are \SI{24}{\centi\metre\per\second} and
\SI{35}{\centi\metre\per\second} 
for the first and second sequence, respectively.

The choice of exposure time for run B was informed by a preliminary analysis
of run A, which showed two possible solutions for the frequency of maximum
oscillation. We therefore opted for faster sampling rate in run B while still 
making sure to reach
the required radial velocity precision for the Rossiter-McLaughlin effect. 

\subsection{Detection of solar-like oscillations in \espresso{} data}
\label{section:oscillations}

\begin{figure}
    \centering
    \includegraphics[width=\linewidth]{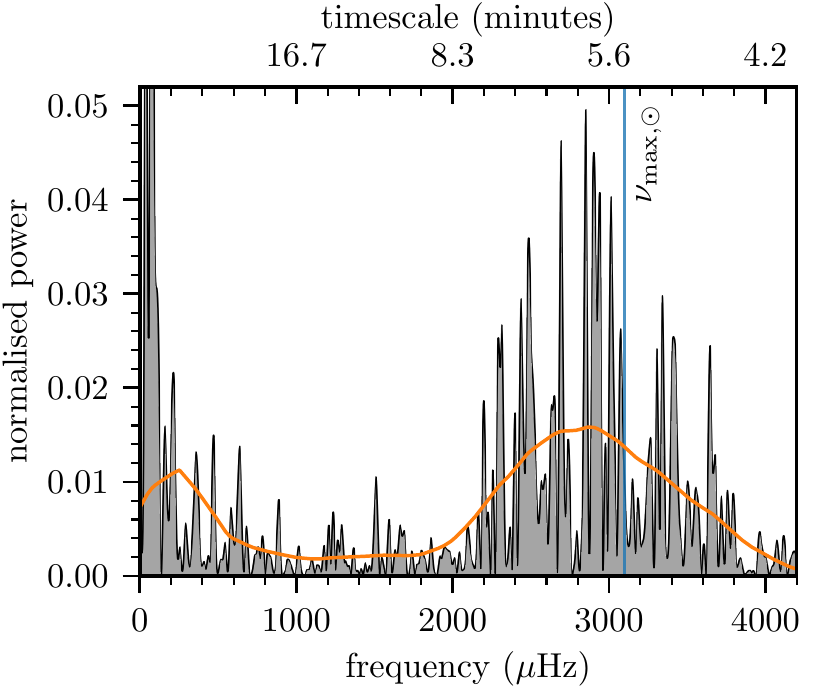}
    \caption{Lomb-Scargle periodogram of the combined \espresso{} radial velocity
        data. The orange curve shows the heavily smoothed spectrum and
    reveals an estimate of $\vmax{} \simeq \SI{2900}{\micro\hertz}$. 
    In blue we denote the frequency of maximum oscillation for
    the Sun for comparison.}
    \label{fig:power_spectrum}
\end{figure}

Our initial approach to obtain the spin--orbit angle of \pimenc{} is to fit the Rossiter-McLaughlin effect from the \espresso{} radial velocity timeseries. The expected amplitude of the Rossiter-McLaughlin effect for this system (assuming $\lambda \sim \SI{0}{\degree}$) is \SI{\sim 50}{\centi\metre\per\second}. In this section, we shall see that the radial velocity timeseries is dominated by variability due to oscillations with amplitudes of a few \si{\metre\per\second}, which makes it challenging to extract the signal of the Rossiter-McLaughlin effect. Before attempting to detect the transiting planet, we first need to characterise the oscillation signal.

Fig.~\ref{fig:power_spectrum} shows the frequency power spectrum of the
\espresso{} radial velocity
residuals. Here, we joined the two segments of data together, and then computed
the spectrum using a Lomb-Scargle periodogram \citep{press1989}, sampled at a
frequency resolution  that corresponds to the inverse of the length of the
combined dataset. The spectrum shows a clear excess of power due to solar-like
oscillations (\emph{p}-modes), centred around a frequency of maximum oscillations power
of $\nu_{\rm max} \simeq \SI{2800}{\micro\hertz}$. Note that the approach of joining
together the two Doppler timeseries does not affect the detectability of the
oscillations, because the damping times of the detectable modes are much shorter
(of the order of a few days) than the gap between the two sets of data.
 
Owing to the very short duration of the combined dataset, it is not possible to
resolve the individual overtones of the oscillation spectrum. The excess power
due to the oscillations may as such be modelled to first order in these data as
a Gaussian or Lorentzian in frequency (here we choose the latter; see, e.g., \citealt{farr2018}). Note the orange line in
Fig.~\ref{fig:power_spectrum} shows the result of
smoothing the spectrum with a double-boxcar filter of width
$\SI{735}{\micro\hertz}$.

We also searched the combined \tess{} residual light curve (after removing the transit and Gaussian process signal) for solar-like oscillations, but
were unable to uncover evidence of detectable modes. However, as we shall see in the next section, the high S/N \espresso{} data do show clearly detectable oscillations in Doppler velocity.


\subsection{Gaussian process modelling of asteroseismic signal and Rossiter-McLaughlin effect}
\label{section:rm_timeseries}

\begin{figure*}
    \centering
    \includegraphics[]{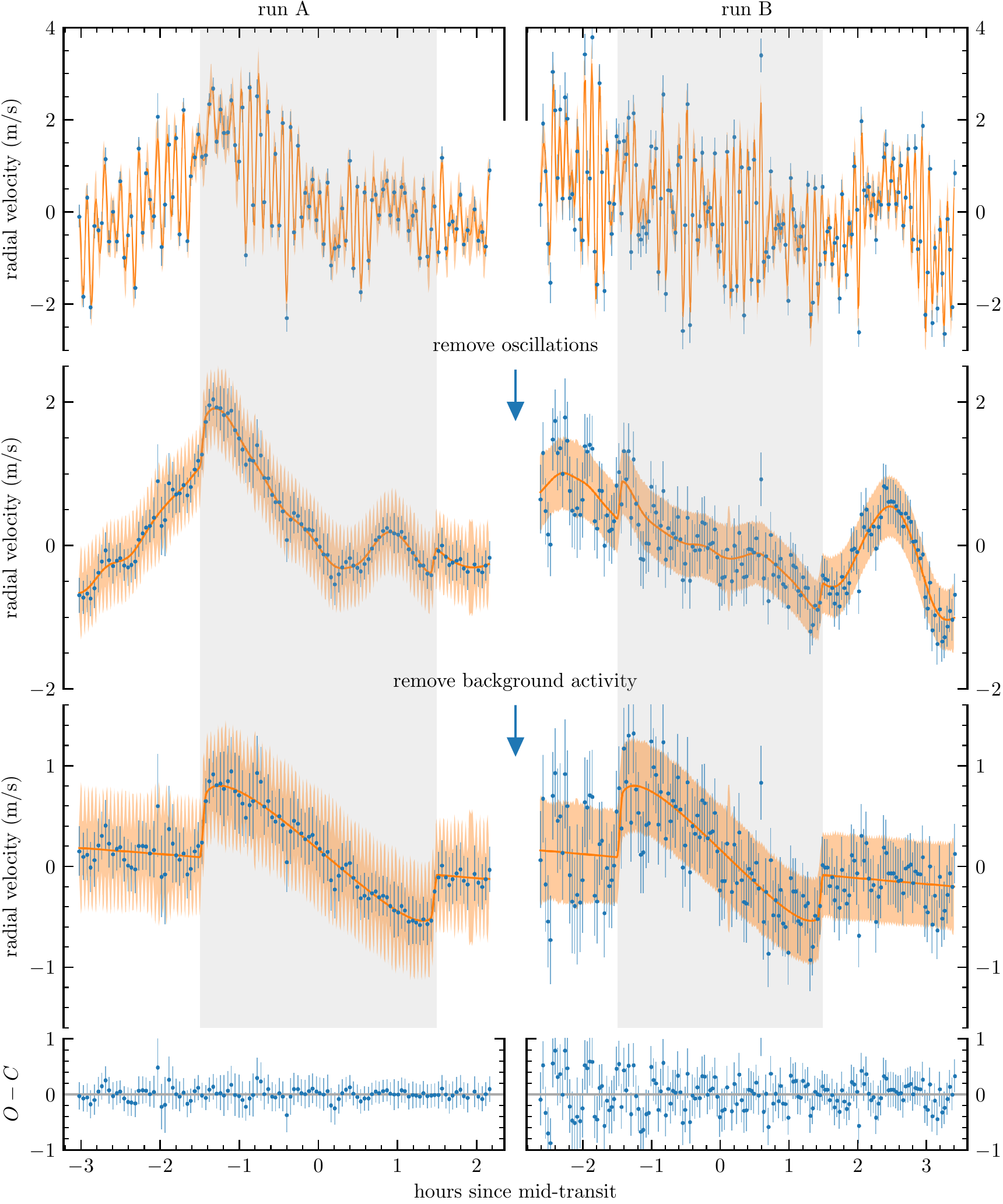}
    \caption{\emph{Top:} \espresso{} data from run A (\emph{left}) and run B (\emph{right}).
    The orange line is the best fit from our combined Rossiter-McLaughlin model
    and Gaussian process (GP) model (oscillations and ``background'' variability). The shaded area is the $68\%$
($1\sigma$) credible interval of the fit. \emph{Upper middle:} The data after
removing the contribution from oscillations in our GP model. \emph{Lower middle:}
The data after further removing the contribution from the longer-term
variability in our GP
model. \emph{Bottom:} Residuals of the fit. The grey, shaded regions denote the transit duration.}
    \label{fig:rm_fit}
\end{figure*}

In this section we build on our analysis from \citet{kunovac-hodzic2019} to try to recover the ``classical'', or velocimetric, Rossiter-McLaughlin effect. 
As is evident from Fig.~\ref{fig:power_spectrum}, 
our data show clear signals from oscillations due to $p$-modes and possibly other
lower frequency stellar activity.
In recent years, Gaussian processes (GPs) have been shown to be robust models
to describe correlated noise, stellar
rotation, or other activity at various timescales (e.g.
\citealt{haywood2014, grunblatt2015, gillen2017, angus2018}). 
Recently, there have also been examples of applying CARMA models
\citep{farr2018} or Gaussian processes \citep{foreman-mackey2017} 
to asteroseismic analyses in the time-domain. This allows for simultaneous transit
or Keplerian modelling, whereas it has traditionally been
done in frequency space. Following \citet{foreman-mackey2017}, 
the oscillations may be modelled 
as a sum of stochastically driven damped simple harmonic oscillators (SHO), 
with power
spectrum $S(\omega)$ given by
\begin{align}
    S(\omega) = \sqrt{\frac{2}{\pi}} \frac{S_0\,\omega_0^4}{(\omega^2 - \omega_0^2)^2 +
    \omega_0^2\omega^2/Q^2},
    \label{eq:sho}
\end{align}
$\omega_0$ is the peak angular frequency, $S_0$ is the maximum power, 
$Q$ is the quality factor that describes the damping timescale.
For large $Q$, Eq.~\ref{eq:sho} approaches a single-peaked Lorentzian function, 
and for modelling oscillations from different modes, $Q$ would describe their 
damping timescale. 
From Section~\ref{section:oscillations} we determined our 
dataset is too short to resolve individual overtones of the power spectrum, thus
we only use a single SHO term (whose power spectrum is described by Eq.~\ref{eq:sho}) to capture the overall shape of the $p$-mode power
excess, and in this case $Q$ 
will effectively describe the width of the power envelope. Similar assumptions
have been made in e.g. \citet{farr2018}. The covariance function of Eq.~\ref{eq:sho} is a type of quasi-periodic function, given by
\begin{align}
     k(\tau) = S_0\,\omega_0\,Q\,\exp{\left(-\frac{\omega_0 \tau}{2Q}\right)} \cos{(\eta \,\omega_0 \,\tau)} + \frac{1}{2\eta Q} \sin{(\eta\,\omega_0 \,\tau)},
     \label{eq:sho_cov}
\end{align}
where $\tau$ is the lag between two measurements in time, and $\eta = \sqrt{|1 - 1/4Q^2|}$. 

In addition to the oscillations, the data show an additional lower frequency
variability component. The variability does not appear
to be periodic, but has characteristics of correlated (red) noise which may originate from stellar surface convection or instrumental/environmental effects. In the case of run A, the background component may be dominated by changes in SNR due to varying observing conditions, see Fig.~\ref{fig:observables}. 
Hereafter we refer to this as
the ``background'' component, and let it encapsulate any variability that is not constrained by the oscillations model or the Rossiter-McLaughlin signal. A similar model to Eq.~\ref{eq:sho} can be used to
describe the background term by fixing $Q = 1/\sqrt{2}$.
In this case, the power spectral density given in Eq.~\ref{eq:sho}
simplifies to
\begin{align}
    S(\omega) = \sqrt{\frac{2}{\pi}} \frac{S_0}{(\omega / \omega_0)^4 + 1},
    \label{eq:granulation}
\end{align}
which describes a Harvey-like model commonly used to describe the background power due to
surface convection (granulation) in asteroseismic and helioseismic analyses
\citep{harvey1985a}, but has also been used to model correlated signals in both radial velocity data and photometry \citep{foreman-mackey2017}. The covariance function for the background model simplies to
\begin{align}
    k(\tau) = S_0\,\omega_0 \exp{\left(-\frac{\omega_0 \tau}{\sqrt{2}}\right)} \cos{\left( \frac{\omega_0 \tau}{\sqrt{2}} - \frac{\pi}{4} \right)}.
    \label{eq:bkg_cov}
\end{align}


We use the
\celerite{} Gaussian processes software \citep{foreman-mackey2017} to model the stellar signals
together with a Rossiter-McLaughlin model. 
We use the simple harmonic oscillator (SHO) kernel within \celerite{}, which is a type of quasi-periodic kernel whose power spectral density and covariance function are described by Eqs.~\ref{eq:sho} and \ref{eq:sho_cov}.%
We fit for
the frequency of maximum oscillation power, $\nu_\mathrm{max} =  \omega_\mathrm{max}/2\pi$, logarithm of the 
amplitude $\ln{S_\mathrm{osc}}$, and power-excess width,
$\ln{Q_\mathrm{osc}}$ to fit the high-frequency 
variations in our data, which was clearly favoured by the Bayesian Information
Criterion (BIC). 
We include an additional SHO term with PSD given by Eq.~\ref{eq:granulation} and covariance function by Eq.~\ref{eq:bkg_cov} for 
the low frequency background component, where $Q$ is fixed to $1/\sqrt{2}$, and is also clearly favoured by the
BIC. Here we fit for the logarithm of the amplitude
$\ln{S_\mathrm{bkg}}$, and angular frequency $\ln{\omega_\mathrm{bkg}}$.
We attempted
to also include a second background term, but found no support for a second
signal from a comparison of the BIC.

The oscillation timescale is only a few times that of our cadence, which can potentially introduce smearing of the oscillation signal. To account for this, we use integrated versions of the \celerite{} kernels that we introduced above (Dan Foreman-Mackey, private comm.). However, we found that this effect did not make a noticeable different on our result. Moreover, inspecting the data in Fig.~\ref{fig:rm_fit}, the noise properties of the two sequences appear qualitatively different. We therefore model the two \espresso{} timeseries with individual Gaussian process kernels since their covariance properties are expected to differ, but share the hyperparameters between them. The priors on the hyperparameters are shown in Table~\ref{table:rv_gp_priors}.

\begin{table}
\renewcommand{\arraystretch}{1.2}
    \centering
    \caption{Priors on the Gaussian process hyperparameters for the radial velocity modelling.}
    \begin{tabular*}{\columnwidth}{@{\extracolsep{\fill}}
            lr}
        \toprule
        \toprule
        {Parameter} & {Prior} \\
        \midrule
        $\ln{S_\text{osc}}$ (\si{\kilo\metre\squared\per\second\squared}) & $\mathcal{U}(-35, 15)$ \\
        $\ln{Q_\mathrm{osc}}$ & $\mathcal{U}(0, 4)$\\
        \vmax{} (\si{\micro\hertz}) & $\mathcal{U}(2100, 4000)$\\
        $\ln{S_\text{bkg}}$ (\si{\kilo\metre\squared\per\second\squared})& $\mathcal{U}(-35, -10)$\\
        $\ln{\omega_\text{bkg}}\ (\text{day}^{-1})$ & $\mathcal{U}(3, 5.75)$\\
        $\ln{\sigma_\text{A}}$ (\si{\kilo\metre\per\second}) & $\mathcal{U}(-15, -4)$\\
        $\ln{\sigma_\text{B}}$ (\si{\kilo\metre\per\second}) & $\mathcal{U}(-15, -4)$ \\
        \bottomrule
    \end{tabular*}
    \label{table:rv_gp_priors}
\end{table}

Our orbital model is computed using 
\ellc{}, where the Rossiter-McLaughlin effect is computed as the flux-weighted radial velocity, as included in the package. 
We vary the
transit parameters 
with Gaussian priors based on the posterior distribution obtained from our 
\tess{} photometric analysis, outlined Appendix~\ref{appendix:tess}, which are listed in Table~\ref{table:results}. 
The radial velocity semi-amplitude is also varied with a Gaussian prior
$K = \SI[separate-uncertainty, multi-part-units=single]{1.5 \pm 0.2}{\metre\per\second}$, according to the most recent radial velocity orbit analysis in \citet{damasso2020}, which is in agreement with 
\citet{huang2018} and \citet{gandolfi2018}.
For the Rossiter-McLaughlin effect we fit for the projected rotation, $\vsini{}$,
of the star, and the spin-orbit angle, $\lambda$, and restrict their values to $\vsini{} \sim \mathcal{U}(0,20)\,\si{\kilo\metre\per\second}$, $\lambda \sim \mathcal{U}(\SI{-180}{\degree}, \SI{180}{\degree})$.

The free parameters in our joint Gaussian process and Rossiter-McLaughlin model are sampled using the \emcee{} package \citep{foreman-mackey2013}. Given the relatively large number of parameters, we launch 400 ``walkers'' that we run for \SI{\sim 50}{} times the estimated autocorrelation length of the parameters. We discard the samples associated with a burn-in phase, which we determined visually, and then checked for convergence by verifying that all parameters reached $\hat{R} < 1.1$ \citep{gelman2003}. The chains were further thinned by their estimated autocorrelation length before merging them, resulting in \num{>1000} effective samples per parameter. 

As we shall see in Section~\ref{section:results}, the detection of the Rossiter-McLaughlin effect is uncertain using the velocimetric method outlined in this section. Therefore, as a next step, we attempt to directly analyse the line profile distortion due to the planet in the next section.

\subsection{Rossiter-McLaughlin reloaded}

\subsubsection{Retrieving the occulted light}
\label{section:rmreloaded_description}

\begin{figure}
    \centering
    \includegraphics[width=\linewidth]{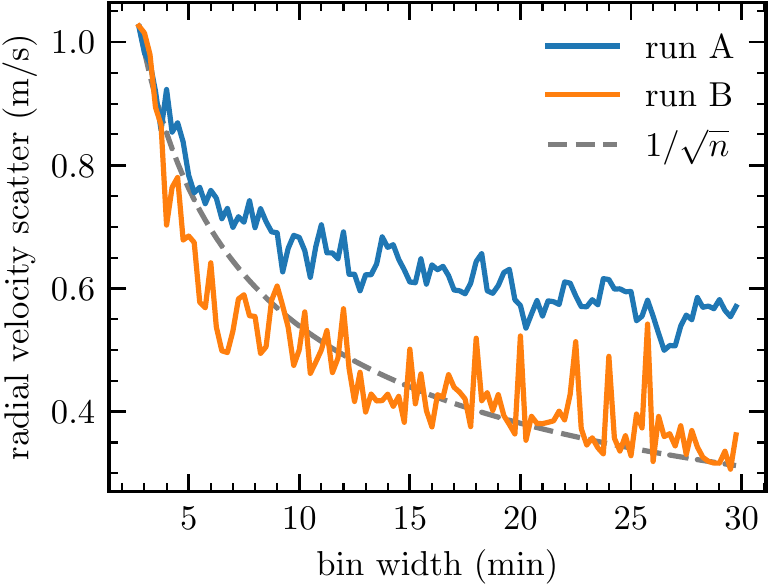}
    \caption{Radial velocity scatter as function of bin width for run A
    (\emph{blue}) and run B (\emph{orange}) The black dashed line denotes the
    expectation from white noise scaling.}
    \label{fig:rms_bin}
\end{figure}

\begin{figure}
    \centering

    \includegraphics[width=\linewidth]{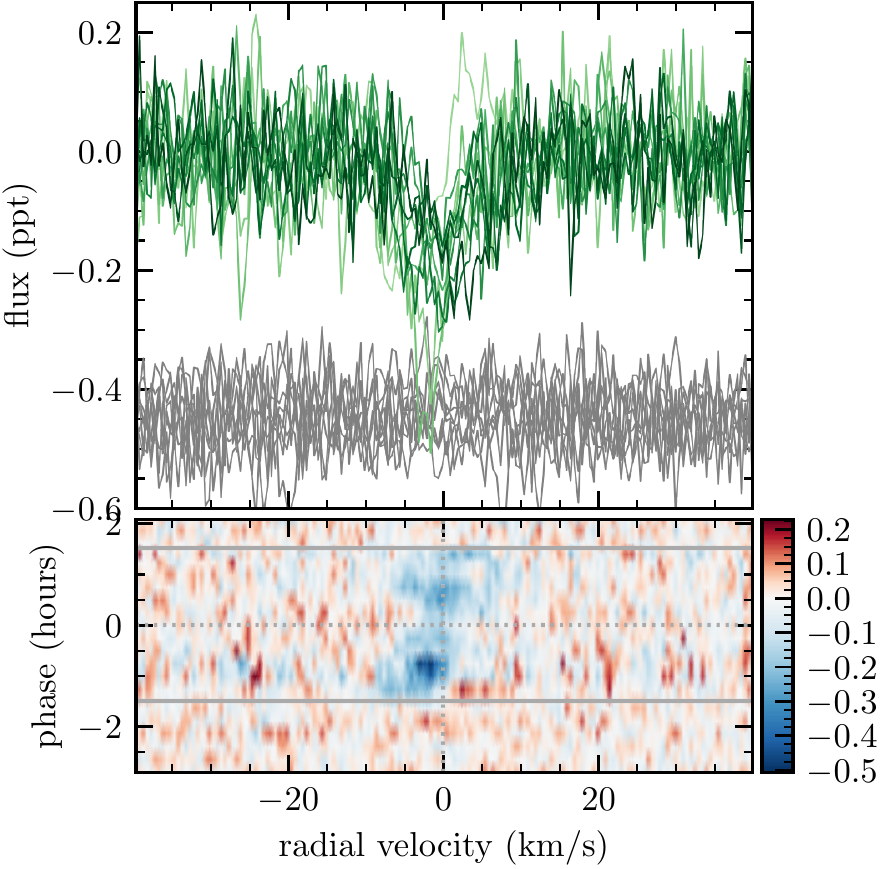}
    \includegraphics[width=\linewidth]{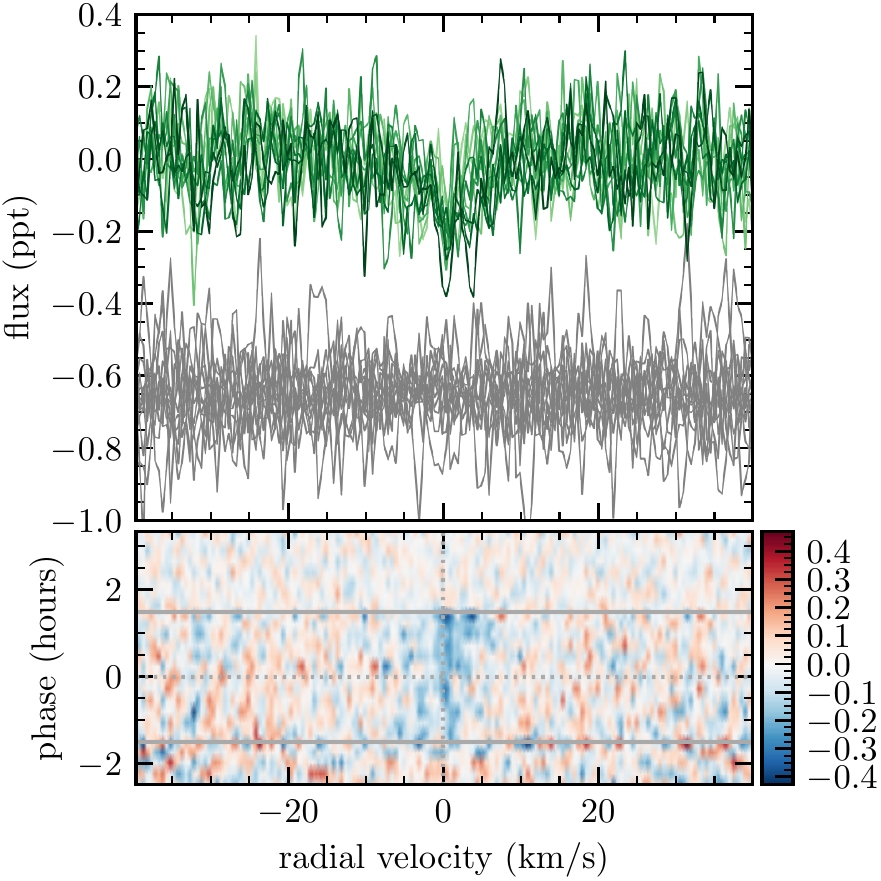}
    \caption{CCF residuals and planet traces for run A (\emph{top}) and run B (\emph{bottom}). In
    each figure, the upper panel shows the residual profiles obtained from subtracting the
binned CCFs from the out of transit reference CCF (\ccfout). The green colours denote in-transit data, getting darker with time. The grey residuals are binned CCFs outside of transit. In the lower panels, the trace of the planet across the
stellar disc is shown, where the grey solid lines denote the transit ingress
and egress, while the white dotted lines denote the stellar rest velocity and
transit mid-point.}
    \label{fig:residual_and_trace} 
\end{figure}

A relatively recent method termed the reloaded Rossiter-McLaughlin method
\citep{cegla2016,bourrier2017,bourrier2018,ehrenreich2020,kunovachodzic2020} uses the cross-correlation functions (CCF) to retrieve the
light on the stellar disc that is occulted by the planet during transit, and has
been shown to address some issues that may bias measurements of $\lambda$ in
``classical'' Rossiter-McLaughlin analyses, such as the method presented in Section~\ref{section:rm_timeseries}. In
summary, the in-transit CCFs (\ccfin{}) are subtracted from a reference
out-of-transit CCF (\ccfout{}) to retrieve the local stellar CCF behind the
planet.
We refer the reader to \citet{cegla2016} for more
details, and also similar, pioneering methods in \citet{albrecht2007} and 
\citet{colliercameron2010}. In the following we will use the abbreviation
``DI'' to refer to ``disc-integrated'' CCFs, to make clear the distinction
from the ``local'' CCFs. Local CCF refers to the retrieved stellar light behind the planet, i.e. Doppler shadow.

The continuum levels of \espresso{} CCFs are arbitrary due to not being flux-calibrated. We therefore start by normalising the DI CCFs by their individual continuum values. Moreover, we also scale them by a quadratic limb darkened
transit model computed from the parameters derived in Appendix~\ref{appendix:tess}, using the same limb darkening coefficients as in the
\tess{} transit analysis.
The DI CCFs are then shifted to the stellar rest frame by removing
the Keplerian motion of the star due to planet c, using the semi-amplitude as derived from \citet{huang2018,gandolfi2018}. The reflex motion of the star due to planet b is smaller -- about \SI{10}{\centi\metre\per\second} and \SI{20}{\centi\metre\per\second} over the transit duration for each run, respectively -- and is therefore a negligible effect. Moreover, we further shift the out-of-transit data to a common systemic velocity at each night, $\gamma$, determined from a weighted average of the out-of-transit disc-integrated radial velocities. The latter step is done in order to build as close to a true, intrinsic line profile of the star as possible, without being affected by the smearing due to the potential large radial velocity variation outside of transit, as can be seen in Fig.~\ref{fig:rm_fit}. 

The signal that we are trying to extract is a very subtle distortion of the DI line profiles, which is expected to cause a shift in radial velocity of just $\SI{\sim 50}{\centi\metre\per\second}$, or put another way; a change in the CCF shape due to the missing flux that is comparable to the transit depth of \SI{\sim 250}{\ppm}. In order to reach this precision, we are required to bin the in-transit DI CCFs to enhance the signal-to-noise of the occulted light. An additional benefit of the binning is to reduce the impact of stellar oscillations. We show in Fig.~\ref{fig:rms_bin} the radial velocity
scatter for a range of bin widths, demonstrating that the oscillations are effectively
binned down as white noise for specific integration times \citep{chaplin2019}. This is particularly clear when the observing conditions are more stable, such as for run B (Fig.~\ref{fig:observables}). 

While binning the DI CCFs equates to longer exposure time and thus higher SNR, it may also have detrimental effects on the signal we are trying to extract. The reloaded Rossiter-McLaughlin effect relies on isolating the differences in the DI CCF shape between in-transit and out-of-transit observations. The contrast (depth) of the DI CCF seems correlated with the SNR, as a lower SNR means fewer stellar absorption lines are resolved when cross-correlating the spectrum with the stellar template. Therefore if the contrast (or FWHM) of the in-transit DI CCFs differ from the out-of-transit DI CCFs at a comparable level to the expected missing light, the residual CCFs (that is the Doppler shadow) may be affected by this difference and thus bias the measurement of its measured radial velocity. This is particularly an issue for run A, where in Fig.~\ref{fig:observables} we show that the contrast during transit varies over a range of \SI{\sim 500}{\ppm}, which is twice the expected signal of the Doppler shadow. In order to attempt to reduce the impact of this effect, we found it best to choose a bin width large enough to effectively reduce the stellar oscillations (Fig.~\ref{fig:rms_bin}) and obtain a high enough SNR, but such that it still bins together DI CCFs with similar contrast. We found that a bin width of 15 minutes was a good compromise taking the above considerations into account, while retaining sampling.

Given the variation in contrast, we found it best to also create master out-of-transit DI CCFs that have characteristics that are as close as possible to the values of the binned in-transit DI CCFs to isolate the Doppler shadow. For each \espresso{} sequence, we create three master out-of-transit CCFs that we refer to as \emph{top}, \emph{middle}, and \emph{bottom}. The top and bottom master CCFs are weighted averages of a combination of the lowest and highest SNR DI CCFs, respectively. The middle master CCF is created from a combination of DI CCFs such that it falls close to the middle of the two. For each binned in-transit DI CCF we compare its contrast with that of the three master CCFs to determine which one it is closest to. From the selected master CCF we subtract the binned in-transit DI CCF to retrieve the planet-occulted residual CCFs (Doppler shadow). The residual profiles are shown in Fig.~\ref{fig:residual_and_trace}, where the Doppler shadow is visible in both runs. The planet appears to be progressively moving from the blueshifted hemisphere to the redshifted hemisphere, indicative of a prograde orbit.

\subsubsection{Retrieving the surface velocities}
\label{section:retrieve_surface_velocity}
We fit Gaussian profiles to the residual line profiles to determine the radial velocity of the surface of the star where it is occulted by the planet. In order to obtain realistic uncertainities and avoid fitting spurious signals, we use a Markov chain Monte Carlo (MCMC) sampling method to explore the full posterior distribution and propagate uncertainties in the nuisance parameters to the final radial velocity. We create our Gaussian model using \pymcthree{} \citep{salvatier2016}, and vary the line centre, $\mu$; width, $\sigma$; contrast, $A$; and continuum level, $c$. Additionally, we also freely fit for the CCF error, $\epsilon$. We use the No-U-Turn (NUTS) sampler \citep{hoffman2014} to sample the parameters from their posterior distribution. Following the procedure in \citet{kunovachodzic2020}, we use wide priors on our parameters related (typically related to the radial velocity grid or flux range range considered), aimed at returning a conservative estimate of the radial velocity in case the local line centre is poorly resolved. We use a half-normal prior on $A$  $\sigma$, and $\epsilon$ as they are restricted to positive values, and a normal prior on $\mu$ and $c$. The priors are shown in Table~\ref{table:pymc3_priors}.

\begin{table}
\renewcommand{\arraystretch}{1.2}
    \centering
    \caption{Priors on the Gaussian model parameters for modelling the residual CCFs. $y$ refers to the residual CCF flux.}
    \begin{tabular*}{\columnwidth}{@{\extracolsep{\fill}}
            lr}
        \toprule
        \toprule
        {Parameter} & {Prior} \\
        \midrule
        $A$  & $\mathcal{N}(0, \text{range}\,y)$ \\
        $\sigma$ (\si{\kilo\metre\per\second}) & $\mathcal{N}(0, 10)$\\
        $\epsilon$ & $\mathcal{N}(0, \text{sd.}\,y)$\\
        $\mu$ (\si{\kilo\metre\per\second})& $\mathcal{N}(0, 40)$\\
        $c$ & $\mathcal{N}(0, \text{sd.}\,y)$\\
        \bottomrule
    \end{tabular*}
    \label{table:pymc3_priors}
\end{table}
From rotational broadening of high-resolution spectra the projected rotation of \pimen{} is estimated to be \SI{\sim3}{\kilo\metre\per\second} \citep{valenti2005}. For physical reasons we therefore further restrict the value of $\mu$ to $[-5, 5]\,\si{\kilo\metre\per\second}$ to avoid fitting spurious correlations that are sometimes seen outside the line core. For each residual CCF we run two individual chains for \num{5000} tuning steps and \num{1000} production steps. This typically results in \num{>500} effective samples per parameter that are well mixed, with $\hat{R} < 1.001$ for all parameters. 
The fits to the individual residual line profiles are shown in Figs.~\ref{fig:ccfpo1}-\ref{fig:ccfpo6}, and the derived radial velocities are shown in Fig.~\ref{fig:local_rv_fit}. Moreover, we show how the derived residual CCF contrast and FWHM compare between the two sequences in Fig.~\ref{fig:local_fwhm_contrast}.

\subsubsection{Local surface velocity modelling}

\begin{figure*}
    \centering
    \includegraphics{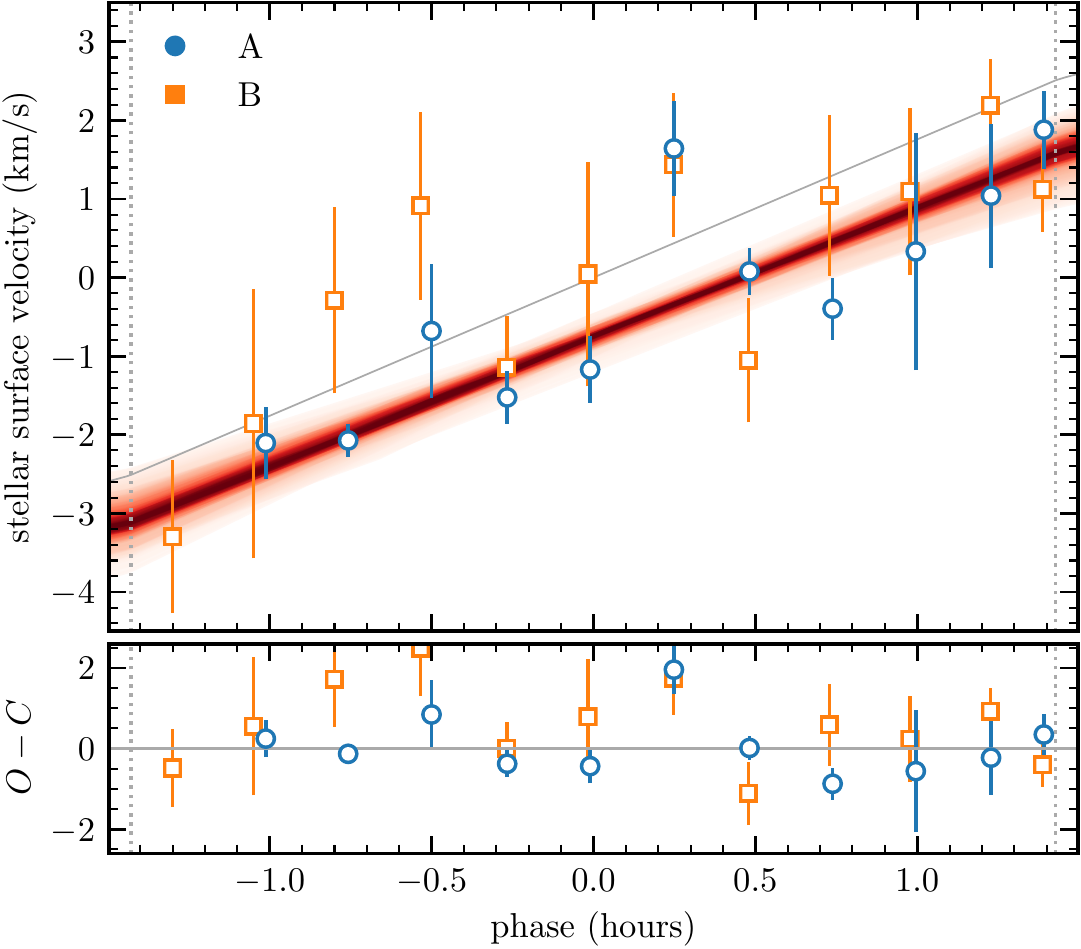}
    \caption{\emph{Upper panel:} The local surface velocities obtained from fitting Gaussian
profiles to the in-transit residual CCFs, shown as blue circles and orange squares for runs A and B, respectively. The red shaded area denote the 50th -- 99th percentiles of the models generated from the posterior distribution of the fit. As a reference, the grey line denotes the $\lambda = \SI{0}{\degree}$ line, which shows that alignment is firmly rejected. \emph{Lower panel:} The residuals from the best-fitting model.}
    \label{fig:local_rv_fit}
\end{figure*}

We model the surface velocities following the semi-analytical model in \citep{cegla2016}. We create a $51 \times 51$ grid that spans the size of the planet, co-moving with its centre. At every observation we compute the brightness-weighted rotational velocity by summing the cells on the stellar disc that are occulted by the planet. We assume the star follows a quadratic limb darkening law with coefficients as reported in the \tess{} analysis. 
We further assume the star follows rigid body rotation, as the precision of our data is not good enough to pick up latitudinal differential rotation. Similarly, we also neglect velocity contributions due to  centre-to-limb convective effects. In this case, the theoretical surface velocity depends on the projected rotational velocity, \vsini{}; projected spin--orbit angle, $\lambda$; the stellar radius scaled by the planet distance, $R_\star/a$; and orbital inclination, $i_\mathrm{p}$. We vary $R_\star/a$ and $i_\mathrm{p}$ within their Gaussian uncertainties determined from the \tess{} transit analysis (Appendix~\ref{appendix:tess}, Table~\ref{table:results}). We let $\vsini{} \sim \mathcal{U}(0, 10)\,\si{\kilo\metre\per\second}$, and $\lambda \sim \mathcal{U}(\SI{-180}{\degree}, \SI{180}{\degree})$. 
We sample these four parameters using the MCMC sampler as implemented in \emcee{} \citep{foreman-mackey2013}, running 200 ``walkers'' for \num{\sim 50} times the estimated autocorrelation length of the parameters. The burn-in phase is discarded by visual inspection of the timeseries of the chains, and we check that we have reached the target posterior distribution by verifying that $\hat{R} < 1.1$ for our parameters \citep{gelman2003}. We further thin the chains by the estimated autocorrelation length, and reach \num{>1000} effective samples per parameter. 
The $50\textsuperscript{th}$ to $99\textsuperscript{th}$ percentile of the models conditioned on the data are shown in Fig.~\ref{fig:local_rv_fit}. The Gaussian fit to the first residual CCF in run A (Appendix~\ref{appendix:ccfpo}) shows signs of a bad comparison with its out-of-transit master CCF due to the rise in flux redwards of the line centre. The line centre is measured at \SI{-4.8}{\kilo\metre\per\second}, which is off the scale in Fig.~\ref{fig:local_rv_fit}. We therefore remove the first data point from run A from our fit to being an outlier. We do however verify that our results do not change with its inclusion. 

\section{Results}
\label{section:results}

We summarise our findings from the \tess{} transit analysis, \espresso{} radial velocity modelling, and reloaded Rossiter-McLaughlin analysis in Table~\ref{table:results}. 
We also report the fitted values for the various Gaussian process nuisance parameters from the photometric and radial velocity analysis in Table~\ref{table:results_hyperparameters}.
In the following, we report our main findings from the analyses presented in this paper.

\subsection{Updated planet radius}


\begin{figure}
    \centering
    \includegraphics[width=\linewidth]{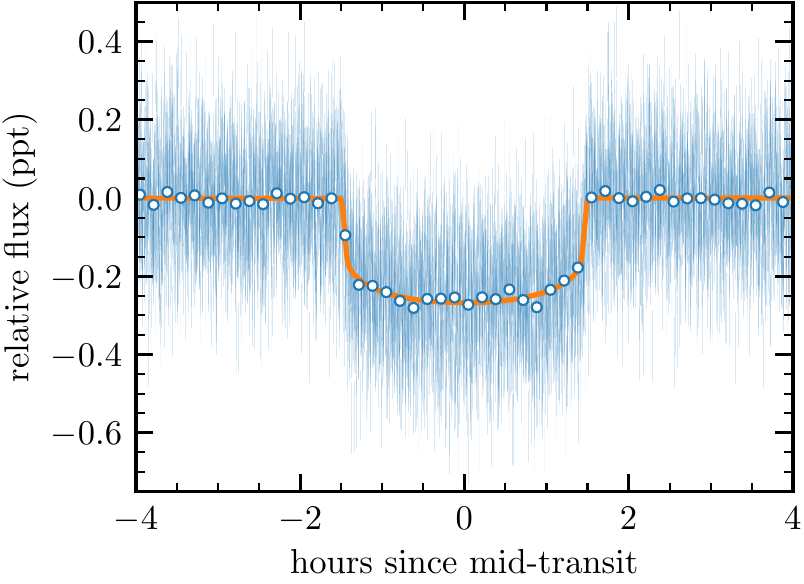}
    \caption{\tess{} data folded on the transit period after removing the best-fit
    Gaussian process model (\emph{blue points}), and binned to 10 minute
averages (\emph{blue/white points}). We also show the best-fit transit model
(\emph{orange line}).}
    \label{fig:pimen_transit}
\end{figure}

\begin{figure}
    \centering
    \includegraphics[width=\linewidth]{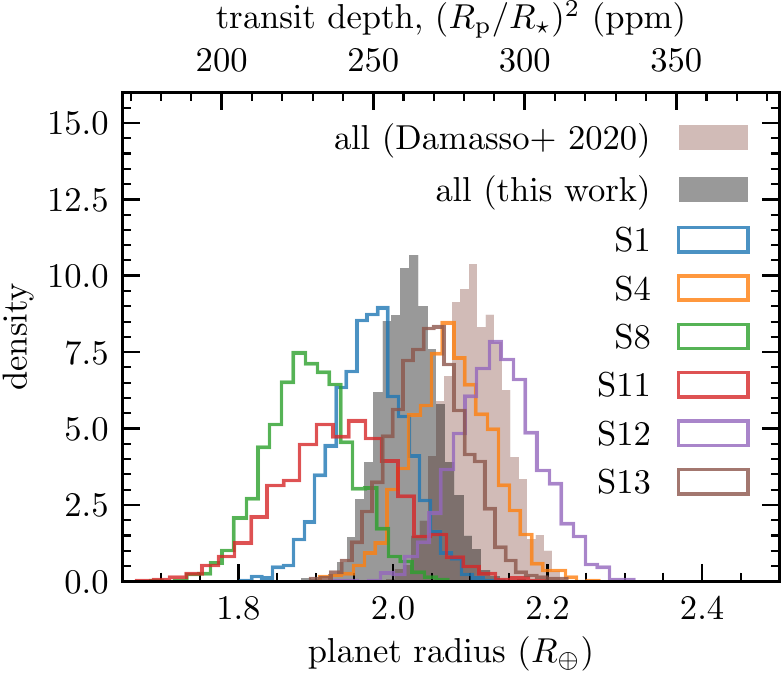}
    \caption{Comparison of the transit depth and planet radius posterior distributions from the MCMC
    sampling from each sector, and the full combined fit using all sectors
    (\emph{gray, filled}) using a transit and Gaussian process model. We also show the posterior distribution from \citet{damasso2020} (\emph{pink, filled}) for comparison. These results are consistent with the previously reported values from \citet{huang2018} and \citet{gandolfi2018}.}
    \label{fig:transit_depths}
\end{figure}

We show the Gaussian process-``detrended" \pimenc{} transit light curve in Fig.~\ref{fig:pimen_transit}, which is based on a combined fit of the six available \tess{} sectors in Cycle 1. 
In Fig.~\ref{fig:transit_depths} we show the transit depths
from fits to individual \tess{} sectors, as well as the combined fit in grey. For
comparison, we include the transit depth posteriors from \citet{damasso2020}, who also modelled the full \tess{} Cycle 1 data. Our results are consistent with the  discovery papers of \citet{huang2018} and \citet{gandolfi2018}, and roughly consistent  (\SIrange{1}{2}{}$\,\sigma$) to \citet{damasso2020}.
The
individual sectors show some spread in their distributions, and although most sectors
agree to about $1\sigma$, Sectors 11 and 8 are \SIrange{2}{3}{\sigma}
discrepant from Sector 12. We attribute these differences to different timescales of correlated noise that is fit to each sector. In presence of both high- and low-frequency variation in the light curve, the Gaussian process will tend to fit the high-frequency component, which may be on a similar timescale as the transit duration, and in turn affect the derived transit depth. Given the new transit depth, we obtain a radius of \pimenc{}, $R_\text{p} =
2.0189^{+0.047}_{-0.045}\,\si{\Rearth}$.

\subsection{Contraints on asteroseismology and the ``classical" Rossiter-McLaughlin effect from radial velocity}

Our Gaussian process model for stellar oscillations, variability, and Rossiter-McLaughlin effect yields a frequency of maximum oscillation power, $\vmax =
2771^{+65}_{-60}\,\mu\mathrm{Hz}$,
which translates to an oscillation period  
of ${\sim}6$ minutes. The longer timescale variability, $\omega_\text{bkg}$, is found to be 
${\sim}2.5$ hours. From Fig.~\ref{fig:power_spectrum} we estimated $\vmax
\simeq \SI{2900}{\micro\hertz}$, but note that the visual estimate is skewed 
towards higher frequencies since the Nyquist frequency of run A is
${\sim}\SI{3000}{\micro\hertz}$ and will thus have increased power close to this
frequency. The Nyquist frequency for run B is however
${\sim}\SI{4000}{\micro\hertz}$ and is thus sampled at a fast enough rate to 
resolve
the true frequency.


Although we do not detect \dnu{} from the \espresso{} analysis, an estimate of the surface gravity can be
obtained from \vmax{} alone through the scaling relation (see
\citealt{chaplin2011} and references therein)
\begin{align}
    \vmax \sim \SI{3090}{\micro\hertz} \frac{g}{g_\odot}
    \sqrt{\frac{T_{\mathrm{eff},\odot}}{T_\mathrm{eff}}}.
    \label{eq:numax_logg}
\end{align}
This gives $\log{g} = 4.4018 \pm 0.0093$ and is consistent within
$1\sigma$ to reported values in the discovery papers as obtained from
spectral analysis in \citet{ghezzi2010}, but more than a factor 3 improvement 
on the uncertainty.

The posterior distribution of \vsini{} shows a \SIrange{2}{3}{}$\sigma$ detection relative to 0, and a spin--orbit angle that suggests misalignment at $1\sigma$. 
Based on the Bayesian Information Criterion (BIC), the 
Rossiter-McLaughlin model is not necessarily favoured, as the Gaussian process
asteroseismic model can easily account for the expected \SI{\sim50}{\centi\metre\per\second} variation. We test this by re-fitting the data without a Rossiter-McLaughlin model. 
We therefore turned to the reloaded Rossiter-McLaughlin method, where we can analyse the line profile distortions directly to detect the spectroscopic transit and measure $\lambda$.

\subsection
[Detection of the Doppler shadow of pi Men c]
{Detection of the Doppler shadow of $\bm{\pi}$ Men\,c}
\label{section:result_rmreloaded}

The stellar surface velocities over the transit duration are plotted in Fig.~\ref{fig:local_rv_fit}.
The two sequences follow the same overall trend, namely a positive
slope for the duration of the transit as the planet moves from blueshifted
to redshifted areas on the disc of the rotating star. This signature is expected 
for a prograde orbit, indicating the Rossiter-McLaughlin effect is detected. The combined fit to both sequences give a projected rotational velocity $\vsini{} = \SI{3.16 \pm 0.27}{\kilo\metre\per\second}$. This value is consistent with the measurements obtained from spectroscopic line broadening in other works \citep{damasso2020}.
The offset of the model from the velocity zero-point at mid-transit determines the projected spin--orbit angle, $\lambda$. We find that the orbit is misaligned  $\ang[angle-symbol-over-decimal]{-24.0} \pm \ang[angle-symbol-over-decimal]{4.1}$ with the stellar spin-axis. The reduced $\chi^2$ ($\chi_\nu^2$) is 1.73 for the joint dataset.

We also fit the two sequences separately to check their consistency. For run A we obtain $\vsini{} = 3.13^{+0.31}_{-0.32}\,\si{\kilo\metre\per\second}$, $\lambda = -27.2^{+4.8}_{-5.3}\,$deg ($\chi_\nu^2 = 2.19$); and for run B we get $\vsini{} = 2.53^{+0.68}_{-0.78}\,\si{\kilo\metre\per\second}$, $\lambda = -8.0^{+12.9}_{-11.6}\,$deg ($\chi_\nu^2 = 1.53$). The joint fit is primarily driven by run A due to its higher SNR data, but \vsini{} for run B is consistent with A to $1\sigma$, and $\lambda$ to within $2\sigma$. The lower value of $\lambda$ for run B is driven by the first four points (in particular the $3\textsuperscript{rd}$ and $4\textsuperscript{th}$), which are observed at low SNR compared to the second half of the transit. Fitting run B without these data gives $\vsini{} = \SI{3.3\pm0.8}{\kilo\metre\per\second}$ and $\lambda = \ang[angle-symbol-over-decimal]{-17} \pm \ang[angle-symbol-over-decimal]{8}$, which is more in line with run A. In Section~\ref{section:discussion_master_choice} we discuss how the choice of master CCF may affect the parameter retrieval.

Using \emph{Hipparcos} data of \pimen{}, \citet{zurlo2018} found a \SI{18.3 \pm 1.0}{\day} periodic signal from a periodogram analysis that they interpreted as stellar rotation. Indeed, the \tess{} Sector 1 light curve show some signs of a \SI{\sim20}{\day} signal, which seem to have disappeared in the three month gap until \pimen{} was observed in the next sector. We can combine the measurement of $P_\text{rot}$ with $R_\star$ to derive the rotational velocity,
$v = 2\pi R_\star/P_\text{rot}$. We can then randomly sample from the distribution of $v$, and the stellar inclination, $\sin{\istar{}}$, whose product can be compared to the measured \vsini{} value from the Rossiter-McLaughlin analysis to retrieve the probability distribution of \istar{} \citep{masuda2020}. We carry out this exercise, but instead sample uniformly in $\cos{\istar{}}$, and find that $\istar{} = \ang[angle-symbol-over-decimal]{90.1} \pm \ang[angle-symbol-over-decimal]{17.3}$. In turn, this allows us to derive the true obliquity, $\psi = 26.9^{+5.8}_{-4.7}\,$deg.

\section{Discussion}
\label{section:discussion}

\subsection
[Robustness of results]
{Robustness of $\bm {v_\text{eq}\sin{i_\star}}$ and $\bm{\lambda}$}
\label{section:discussion_master_choice}

We test the robustness of our results using a number of procedures. First, we check whether our results are driven by a few points from run A that have low uncertainties, in particular the points at $-1.0\,$h and $-0.75\,$h and $+0.5\,$h, whose errors might be underestimated considering some of these have a low SNR disc-integrated cross-correlation function (DI CCF). We repeat our fitting procedure by excluding these three points and find $\vsini{} = \SI{2.92 \pm 0.45}{\kilo\metre\per\second}$, $\lambda = \SI[multi-part-units=repeat]{-23 \pm 5}{\degree}$, which is 
consistent to our previous analysis.

Next, we test the sensitivity of \vsini{} and $\lambda$ to our choice of fitting the residual CCF uncertainty, $\epsilon$, to make sure the CCF uncertainties are not underestimated. We re-analyse the Gaussian fits to the residual CCFs from both runs, but this time fix the residual CCF uncertainties to the values provided by the \espresso{} data products. Here we also retrieve consistent results, $\vsini{} = \SI{2.92 \pm 0.15}{\kilo\metre\per\second}$, $\lambda = \SI[multi-part-units=repeat]{-19 \pm 2}{\degree}$,
albeit with smaller uncertainties because the fixed \espresso{} CCF uncertainties are up to \SI{30}{\percent} smaller than our fitted uncertainty, $\epsilon$. We therefore opted for the more conservative approach of fitting $\epsilon$.

We also test how sensitive our results are to the uncertainty in the centre of the line profile. A Gaussian fit to the averaged out-of-transit disc-integrated CCFs gives an uncertainty in the radial velocity centre of \SI{\sim 3}{\centi\metre\per\second}, which is below the \SI{10}{\centi\metre\per\second} instrumental precision of \espresso{}. Never the less, we repeat our analysis by perturbing the line centre velocity by \SI{\pm 3}{\centi\metre\per\second}, and find that our results are not very sensitive to the line centre uncertainty, $\vsini{} = \SI{3.06 \pm 0.25}{\kilo\metre\per\second}$, $\lambda = \SI[multi-part-units=repeat]{-19 \pm 4}{\degree}$ and $\vsini{} = \SI{3.06 \pm 0.25}{\kilo\metre\per\second}$, $\lambda = \SI[multi-part-units=repeat]{-24 \pm 4}{\degree}$ (by adding and subtracting, respectively). The reduced $\chi^2$ does however slightly increase from $1.7$ to $2.3$, which indicates our original solution is optimized.

It is reasonable to assume that the particular selection of spectra to build the master out-of-transit CCF in the reloaded Rossiter-McLaughlin method can impact the results. Indeed, we experimented with several ways of building the reference CCFs. Run A was particularly sensitive to this choice due to the variation in CCF shape during transit from variable observing conditions. Doing the standard approach, i.e. averaging all out-of-transit CCFs into one master CCF that is applied equally to all the binned in-transit points led to significant correlations of the retrieved surface velocities with the DI CCF contrasts. Typically the retrieved parameters using this approach would be $\vsini{} \sim \SI{5}{\kilo\metre\per\second}$ and $\lambda \sim \SI{-50}{\degree}$ for $\chi_\nu^2 >> 1$. Besides the poor fit, the \vsini{} measurement was inconsistent with that determined from rotational broadening, and neither parameter was supported by the retrieved values for run B. We also attempted to build custom master CCFs for each binned in-transit CCF by selecting out-of-transit CCFs that has similar depths, within a pre-determined range. However, this approach led to the in-transit CCFs at the lowest and highest SNR/contrast (beginning and end of transit, Fig.~\ref{fig:observables}) to have very few matches, which led to a low SNR master CCF. This again lead to some correlations in the retrieved surface velocities, and a solution that is inconsistent with the priors we placed on $R_\star/a$ and $i_\text{p}$. Out of the many methods we experimented with, the method we outlined in Section~\ref{section:rmreloaded_description} was the only one that led to 
few correlations in the surface velocities, gave \vsini{} consistent with the expectation from spectroscopic broadening, and overall less complex.

While the SNR of run B is overall lower than that of run A due to faster sampling and thus more time spent reading the CCD, the observing conditions on that night are much more stable, and gradually improve for the entire duration of the transit. This leads to a slope in the DI CCF contrast of the in-transit data that is easier to deal with. Ultimately we chose the same strategy for creating master CCFs as for run A for consistency. Overall, we found that the retrieved \vsini{} and $\lambda$ for run B are less sensitive to the choice of master CCF. For example, the DI CCFs after the transit of run B have higher SNR than the pre-transit data. A more conservative -- and less complex -- approach for run B would be to use all the post-transit CCFs as reference. This gives $\vsini{} \approx \SI{3.5}{\kilo\metre\per\second}$, $\lambda \approx \SI{-30}{\degree}$. This is more consistent with the result from run A. However, due to lower SNR at the beginning of transit, this method leads to three of the residual profiles being flat. Similarly, we can also create custom master CCFs for each binned in-transit CCF as we outlined for run A in the previous paragraph. This method produces similar \vsini{}, $\lambda \approx \SI{-17}{\degree}$.

Ultimately, the method of creating master CCFs should be determined based on the specific characteristics of each data set, which we have shown is different between run A and B. Adopting a less complex and conservative approach for run B (but sacrificing consistency) we obtain fully consistent results with run A, namely that the orbit of \pimenc{} is misaligned with the stellar spin by about \SI{-30}{\degree}.


\subsection{Dynamical history}

The formation of super-Earth and sub-Neptune multiplanetary systems is a subject
of intense debate
\citep{raymond2018,schoonenberg2019,coleman2019,lee2014,mohanty2018, wu2019}.
To this day it remains unclear whether super-Earths and sub-Neptunes form within
the water iceline, or whether they can form beyond. Either, or both might be
true. Detections of super-Earths in circumbinary configurations \citep[e.g.][]{orosz2019,kostov2020} point out that super-Earths can likely form at large distances before migrating inwards \citep{martin2018,pierens2020}.

When proposing the observations reported here for DDT, we had initially speculated that the architecture of the \pimen{} system was reminiscent of past dynamical interactions, and wanted to verify this with Rossiter-McLaughlin timeseries. These suspicions are now confirmed by the inclined orbital plane of the inner planet \pimenc{}, but also thanks to three recent analyses that appeared while we were finalising our own paper. \citet{xuan2020}, \citet{damasso2020}, and \citet{derosa2020} combined radial-velocities with {\it Gaia} and {\it Hipparcos} astrometry, and measured that \pimenb{} is inclined with $i_b = \ang[angle-symbol-over-decimal]{45.8}^{+1.4^\circ}_{-1.1^\circ} $ \citep{damasso2020}, which also shows a large mutual inclination between \pimenb{} and \pimenc{}.

Together this provides
evidence that super-Earths can likely form beyond the iceline around single stars.
On its own, the inclination of \pimenb{} did not imply much for \pimenc{}. However, a natural cause for \pimenc{}'s own orbital inclination is an exchange of angular momentum between the outer and
inner planetary orbit (e.g \citealt{wu2011, matsumura2010}). Currently the
two planets are likely too distant from one another to allow such an
exchange. To allow raised inclinations, planet b and c would have needed to be closer. Due to their mass and angular momentum ratios, it is more likely that planet c moved inwards than for planet b to move outwards, this would imply that planet c might have formed near or beyond the iceline. Detailed N-body numerical simulations will be necessary to explore this scenario further.

\section{Conclusion}
\label{section:conclusion}
In this paper we have presented two high-resolution spectroscopic transits of the super-Earth \pimenc{} that we observed with \espresso{}. The two timeseries show a rich concoction of signals of various origins. We perform an asteroseismic analysis on the radial velocity timeseries using Gaussian processes and clearly detect the frequency of maximum oscillation, \vmax{}. We are not sensitive to the measurement of the large frequency separation, \dnu{}, and can therefore not directly measure the stellar mass and radius, but are able to obtain a much tighter constraint on $\log{g}$ than from spectroscopy.

We performed a transit analysis on all the available data of \pimen{} from \tess{} Cycle 1; a total of six sectors. We revise the transit parameters for \pimenc{} 
and find a radius of $R_\text{p} = 2.019^{+0.047}_{-0.045}\,\si{\Rearth}$, which is roughly consistent with previous analyses in \citealt{damasso2020,gandolfi2018,huang2018}.

We attempted to fit the \espresso{} radial velocity timeseries to isolate ``classical'', i.e. velocimetric, Rossiter-McLaughlin effect, but the radial velocity signal was thwarted by asteroseismic p-modes and a longer term variation that we attribute to a form of stellar granulation. Instead, we turned to the reloaded Rossiter-McLaughlin method to detect the Doppler shadow. Using this method we were able to detect the transit of \pimenc{} in spectroscopy using \espresso{}, and found that its orbit is misaligned by $\ang[angle-symbol-over-decimal]{-24.0} \pm \ang[angle-symbol-over-decimal]{4.1}$ with the stellar spin-axis. This makes \pimenc{} the smallest mass-ratio planet measured with the Rossiter-McLaughlin effect. Combining our results with the recent detection of the inclination of the external planet \citep{xuan2020,damasso2020,derosa2020}, we speculate that \pimenc{} likely formed at a large distance from the star. There it gravitationally interacted with \pimenb{}, an interaction which is still evident from their mutual inclinations, \pimenc{}'s own inclination with respect to the stellar spin axis, and the high eccentricity of \pimenb{}.

This work stands as a testament to a new avenue of scientific investigation, namely the origins of super-Earth and sub-Neptune planets, which is now made possible by the extreme precision of \espresso{}.




\section*{Acknowledgements} 
We would like to thank the referee for their thorough and constructive comments that substantially improved the paper.
Parts of this work was carried out during a Fulbright Fellowship at the University of Chicago, funded by the U.S.--Norway Fulbright Foundation. VKH would like to thank Daniel Fabrycky and David Martin for kindly hosting him at the University of Chicago, and for useful discussions on this work. We would also like to thank Dan Foreman-Mackey for help on integrated \celerite{} kernels and separation of kernel predictions, and Vincent Bourrier for discussions on strategies for mitigating stellar oscillations. VKH is supported by a Birmingham Doctoral Scholarship, and by a studentship from Birmingham's School of Physics and Astronomy. This research received funding from the European Research Council (ERC) under the European Union's Horizon 2020 research and innovation programme (grant agreement n$^\circ$ 803193/BEBOP and 804752/CartographY). HMC acknowledges the financial support of the National Centre for Competence in Research PlanetS supported by the Swiss National Science Foundation (SNSF), and the UK Research Innovation Future Leaders Fellowship (MR/S035214/1). 
This paper includes data collected by the TESS mission. Funding for the TESS mission is provided by the NASA's Science Mission Directorate. 
This research made use of \exoplanet{} \citep{foreman-mackey2019} and its
dependencies \citep{agol2020, astropycollaboration2013, astropycollaboration2018,
foreman-mackey2017, foreman-mackey2018,
luger2019, salvatier2016, thetheanodevelopmentteam2016}.
This research also made use of the the open-source \textsf{python} packages \emcee{} \citep{foreman-mackey2013}, \ellc{} \citep{maxted2016}, \textsf{numpy} \citep{vanderwalt2011}, \textsf{scipy} \citep{jones2002}, and \textsf{matplotlib} \citep{hunter2007a}.




\section*{Data availability}
The spectroscopic data underlying this article are available from the public ESO archive\footnote{\url{www.archive.eso.org}}. The photometric data are publicly available from the Mikulski Archive for Space Telescopes (MAST) portal.\footnote{\url{https://mast.stsci.edu/portal/Mashup/Clients/Mast/Portal.html}}

\begin{table*}
\renewcommand{\arraystretch}{1.2}
    \centering
    \caption{System parameters for \pimen{} (HD 39091) and derived planet parameters from the \espresso{} and \tess{} analysis. $^{\textblue{a}}$Determined from spectral broadening. $^{\textblue{b}}$Fixed. $^{\textblue{c}}$Derived from \vmax{} using Equation~\ref{eq:numax_logg}. 
    $^{\textblue{d}}$Derived from the combination of $P_\text{rot}$, $R_\star$, and \vsini{}.}
    \begin{tabular*}{\linewidth}{@{\extracolsep{\fill}}
            llcc}
        \toprule
        \toprule
        {\textbf{Parameter}} & {\textbf{Description}} & {\textbf{Value}} & {\textbf{Reference/data}}  \\
        \midrule
        \multicolumn{4}{@{}l@{}}{\emph{System information and stellar parameters}} \\
        \midrule
        \emph{Gaia} DR2 ID & -- & 4623036865373793408 & Simbad \\
        \emph{TIC} & \tess{} Input Catalog ID & 261136679 & MAST \\
        $\alpha$ & Right ascension & \ra[angle-symbol-over-decimal]{05;37;09.89} & Simbad \\
        $\delta$ & Declination & \ang[angle-symbol-over-decimal]{-80;28;08.8} & Simbad \\
        $V$ (mag) & Apparent magnitude & 5.67 & Simbad \\
        Distance (pc) & Parallax distance & $18.27 \pm 0.02$ & \text{\cite{gaiacollaboration2018}} \\
        $T_\mathrm{eff}$ (\si{\kelvin}) & Effective temperature & $5998 \pm 62$ & \text{\cite{damasso2020}}\\
        $\log{g}$ (cgs) & Stellar surface gravity & $4.43 \pm 0.1$ & \text{\cite{damasso2020}}\\
        $[\mathrm{Fe/H}]$ (dex) & Stellar metallicity & $0.09 \pm 0.04$ & \text{\cite{damasso2020}}\\
        \vsini{} (\si{\kilo\metre\per\second}) & Projected rotational velocity & $3.34 \pm 0.07^{\textblue{a}}$ &
        \text{\cite{damasso2020}} \\
        $M_\star$ (\si{\Msun}) & Stellar mass &  $1.07 \pm 0.04$ & \text{\cite{damasso2020}}\\
        $R_\star$ (\si{\Rsun}) & Stellar radius & $1.17 \pm 0.02$ & \text{\cite{damasso2020}}\\
        $L_\star$ (\si{\Lsun}) & Stellar luminosiity & $1.44 \pm 0.02$ & \text{\cite{huang2018}} \\
        $\tau_\star$ (Gyr) & Stellar age & $2.98^{+1.4}_{-1.3}$ & \text{\cite{huang2018}}\\
        $P_\text{rot}$ (days) & Photometric rotation period & $18.3 \pm 1.0$ & \text{\cite{zurlo2018}} \\[5pt]

        \multicolumn{4}{@{}l@{}}{\emph{Transit parameters}} \\
        \midrule
        $P$ (days)  & Orbital period & $6.2678399 \pm 0.000011$ & \tess{} \\
        
        $T_0$ ($\si{\bjdutc{}} - \num{2450000}$) & Transit mid-point & 
        $8425.789204^{+0.000279}_{-0.000280}$ & \tess{} \\
        
        $R_\text{p}$ (\si{\Rearth}) & Planet radius & $2.0189^{+0.0465}_{-0.0448}$ & \tess{} \\
        
        $R_\text{p}/R_\star$ & Planet-to-star radius ratio & 
        $0.015844^{+0.000139}_{-0.000136}$ & \tess{} \\
        
        $D_{T_0}$ (ppm) & Transit depth at $T_0$ & $267.59^{+4.16}_{-3.89}$ & \tess{} \\
        
        $R_\star/a$ & Scaled separation &  $0.07978^{+0.00175}_{-0.00172}$ & \tess{} \\
        
        $b$ ($R_\star$) & Impact parameter & $0.6465^{+0.0191}_{-0.0210}$ & \tess{} \\
        
        
        $i_\text{p}$ (\si{\degree}) & Orbital inclination & $87.045^{+0.157}_{-0.156}$ & \tess{} \\
        
        $T_{14}$ (days) & Transit duration between $1\textsuperscript{st}$ and $4\textsuperscript{th}$ contacts & $0.125001^{+0.000484}_{-0.000464}$ & \tess{} \\
        
        $e$ & Eccentricity & 0$^{\textblue{b}}$ & \tess{} \\
        
        $\omega$ (\si{\degree}) & Argument of periastron & 0$^{\textblue{b}}$ & \tess{} \\[5pt]
        
        \multicolumn{4}{@{}l@{}}{\emph{``Classical'' Rossiter-McLaughlin and asteroseismology}} \\
        \midrule
        
        
        $\vsini{}$ (\si{\kilo\metre\per\second}) & Projected rotation velocity & $4.29^{+1.57}_{-1.58}$ & \espresso{} RVs\\
        $\lambda$ (\si{\degree}) & Spin--orbit angle & $-20.4^{+26.4}_{-24.6}$ & \espresso{} RVs\\
        $\vmax{}$ (\si{\micro\hertz}) & Frequency of maximum oscillation & $2771.2^{+65.3}_{-59.9}$ & \espresso{} RVs\\
        $\log{g}$ (cgs) & Stellar surface gravity$^{\textblue{c}}$ & $4.4018 \pm 0.0093$ & \espresso{} RVs \\ [5pt]

        \multicolumn{4}{@{}l@{}}{\emph{Reloaded Rossiter-McLaughlin}} \\
        \midrule
        \vsini{} (\si{\kilo\metre\per\second}) & Projected rotational velocity (run A,B) & $3.16 \pm 0.27$ & \espresso{} CCFs \\
        $\lambda$ (\si{\degree}) & Projected spin--orbit angle (run A,B) & $-24.0 \pm 4.1$ & \espresso{} CCFs \\
        $\istar{}$ (\si{\degree}) & Stellar inclination$^{\textblue{d}}$ & $90.1 \pm 17.3$ & $P_\text{rot}$, $R_\star$, \vsini{}\\
        $\psi$ (\si{\degree}) & 3D spin--orbit angle$^{\textblue{d}}$ & $26.9^{+5.8}_{-4.7}$ & $\lambda$, $i_\text{p}$, $i_\star$ \\
        \bottomrule
    \end{tabular*}
    \label{table:results}
\end{table*}

\bibliographystyle{mnras}
\bibliography{references} 

\begin{thebibliography}{}
\makeatletter
\relax
\def\mn@urlcharsother{\let\do\@makeother \do\$\do\&\do\#\do\^\do\_\do\%\do\~}
\def\mn@doi{\begingroup\mn@urlcharsother \@ifnextchar [ {\mn@doi@}
  {\mn@doi@[]}}
\def\mn@doi@[#1]#2{\def\@tempa{#1}\ifx\@tempa\@empty \href
  {http://dx.doi.org/#2} {doi:#2}\else \href {http://dx.doi.org/#2} {#1}\fi
  \endgroup}
\def\mn@eprint#1#2{\mn@eprint@#1:#2::\@nil}
\def\mn@eprint@arXiv#1{\href {http://arxiv.org/abs/#1} {{\tt arXiv:#1}}}
\def\mn@eprint@dblp#1{\href {http://dblp.uni-trier.de/rec/bibtex/#1.xml}
  {dblp:#1}}
\def\mn@eprint@#1:#2:#3:#4\@nil{\def\@tempa {#1}\def\@tempb {#2}\def\@tempc
  {#3}\ifx \@tempc \@empty \let \@tempc \@tempb \let \@tempb \@tempa \fi \ifx
  \@tempb \@empty \def\@tempb {arXiv}\fi \@ifundefined
  {mn@eprint@\@tempb}{\@tempb:\@tempc}{\expandafter \expandafter \csname
  mn@eprint@\@tempb\endcsname \expandafter{\@tempc}}}

\bibitem[\protect\citeauthoryear{Addison et~al.,}{Addison
  et~al.}{2020}]{addison2020}
Addison B.~C.,  et~al., 2020, arXiv e-prints, 2006, arXiv:2006.13675

\bibitem[\protect\citeauthoryear{Agol, Luger  \& {Foreman-Mackey}}{Agol
  et~al.}{2020}]{agol2020}
Agol E.,  Luger R.,   {Foreman-Mackey} D.,  2020, \mn@doi [The Astronomical
  Journal] {10.3847/1538-3881/ab4fee}, 159, 123

\bibitem[\protect\citeauthoryear{Albrecht, Reffert, Snellen, Quirrenbach  \&
  Mitchell}{Albrecht et~al.}{2007}]{albrecht2007}
Albrecht S.,  Reffert S.,  Snellen I.,  Quirrenbach A.,   Mitchell D.~S.,
  2007, \mn@doi [Astronomy and Astrophysics] {10.1051/0004-6361:20077953}, 474,
  565

\bibitem[\protect\citeauthoryear{Angus, Morton, Aigrain, {Foreman-Mackey}  \&
  Rajpaul}{Angus et~al.}{2018}]{angus2018}
Angus R.,  Morton T.,  Aigrain S.,  {Foreman-Mackey} D.,   Rajpaul V.,  2018,
  \mn@doi [Monthly Notices of the Royal Astronomical Society]
  {10.1093/mnras/stx2109}, 474, 2094

\bibitem[\protect\citeauthoryear{Armitage}{Armitage}{2013}]{armitage2013}
Armitage P.~J.,  2013, Astrophysics of Planet Formation

\bibitem[\protect\citeauthoryear{{Astropy Collaboration} et~al.,}{{Astropy
  Collaboration} et~al.}{2013}]{astropycollaboration2013}
{Astropy Collaboration} et~al., 2013, \mn@doi [Astronomy and Astrophysics]
  {10.1051/0004-6361/201322068}, 558, A33

\bibitem[\protect\citeauthoryear{{Astropy Collaboration} et~al.,}{{Astropy
  Collaboration} et~al.}{2018}]{astropycollaboration2018}
{Astropy Collaboration} et~al., 2018, \mn@doi [The Astronomical Journal]
  {10.3847/1538-3881/aabc4f}, 156, 123

\bibitem[\protect\citeauthoryear{Barclay, Pepper  \& Quintana}{Barclay
  et~al.}{2018}]{barclay2018}
Barclay T.,  Pepper J.,   Quintana E.~V.,  2018, \mn@doi [The Astrophysical
  Journal Supplement Series] {10.3847/1538-4365/aae3e9}, 239, 2

\bibitem[\protect\citeauthoryear{Baruteau et~al.,}{Baruteau
  et~al.}{2014}]{baruteau2014}
Baruteau C.,  et~al., 2014, \mn@doi [Protostars and Planets VI]
  {10.2458/azu_uapress_9780816531240-ch029}, p.~667

\bibitem[\protect\citeauthoryear{Batalha et~al.,}{Batalha
  et~al.}{2013}]{batalha2013}
Batalha N.~M.,  et~al., 2013, \mn@doi [The Astrophysical Journal Supplement
  Series] {10.1088/0067-0049/204/2/24}, 204, 24

\bibitem[\protect\citeauthoryear{Bitsch, Crida, Libert  \& Lega}{Bitsch
  et~al.}{2013}]{bitsch2013}
Bitsch B.,  Crida A.,  Libert A.-S.,   Lega E.,  2013, \mn@doi [Astronomy and
  Astrophysics] {10.1051/0004-6361/201220310}, 555, A124

\bibitem[\protect\citeauthoryear{Bonfils et~al.,}{Bonfils
  et~al.}{2013}]{bonfils2013}
Bonfils X.,  et~al., 2013, \mn@doi [Astronomy and Astrophysics]
  {10.1051/0004-6361/201014704}, 549, A109

\bibitem[\protect\citeauthoryear{Borucki et~al.,}{Borucki
  et~al.}{2010}]{borucki2010}
Borucki W.~J.,  et~al., 2010, \mn@doi [Science] {10.1126/science.1185402}, 327,
  977

\bibitem[\protect\citeauthoryear{Bourrier \& H{\'e}brard}{Bourrier \&
  H{\'e}brard}{2014}]{bourrier2014}
Bourrier V.,  H{\'e}brard G.,  2014, \mn@doi [Astronomy and Astrophysics]
  {10.1051/0004-6361/201424266}, 569, A65

\bibitem[\protect\citeauthoryear{Bourrier, Cegla, Lovis  \&
  Wyttenbach}{Bourrier et~al.}{2017}]{bourrier2017}
Bourrier V.,  Cegla H.~M.,  Lovis C.,   Wyttenbach A.,  2017, \mn@doi
  [Astronomy and Astrophysics] {10.1051/0004-6361/201629973}, 599, A33

\bibitem[\protect\citeauthoryear{Bourrier et~al.,}{Bourrier
  et~al.}{2018}]{bourrier2018}
Bourrier V.,  et~al., 2018, \mn@doi [Nature] {10.1038/nature24677}, 553, 477

\bibitem[\protect\citeauthoryear{Cegla, Lovis, Bourrier, Beeck, Watson  \&
  Pepe}{Cegla et~al.}{2016}]{cegla2016}
Cegla H.~M.,  Lovis C.,  Bourrier V.,  Beeck B.,  Watson C.~A.,   Pepe F.,
  2016, \mn@doi [Astronomy and Astrophysics] {10.1051/0004-6361/201527794},
  588, A127

\bibitem[\protect\citeauthoryear{Chabrier, Johansen, Janson  \&
  Rafikov}{Chabrier et~al.}{2014}]{chabrier2014}
Chabrier G.,  Johansen A.,  Janson M.,   Rafikov R.,  2014, \mn@doi [Protostars
  and Planets VI] {10.2458/azu_uapress_9780816531240-ch027}, pp 619--642

\bibitem[\protect\citeauthoryear{Chaplin et~al.,}{Chaplin
  et~al.}{2011}]{chaplin2011}
Chaplin W.~J.,  et~al., 2011, \mn@doi [The Astrophysical Journal]
  {10.1088/0004-637X/732/1/54}, 732, 54

\bibitem[\protect\citeauthoryear{Chaplin, Cegla, Watson, Davies  \&
  Ball}{Chaplin et~al.}{2019}]{chaplin2019}
Chaplin W.~J.,  Cegla H.~M.,  Watson C.~A.,  Davies G.~R.,   Ball W.~H.,  2019,
  \mn@doi [The Astronomical Journal] {10.3847/1538-3881/ab0c01}, 157, 163

\bibitem[\protect\citeauthoryear{Chatterjee, Ford, Matsumura  \&
  Rasio}{Chatterjee et~al.}{2008}]{chatterjee2008}
Chatterjee S.,  Ford E.~B.,  Matsumura S.,   Rasio F.~A.,  2008, \mn@doi [The
  Astrophysical Journal] {10.1086/590227}, 686, 580

\bibitem[\protect\citeauthoryear{Coleman, Leleu, Alibert  \& Benz}{Coleman
  et~al.}{2019}]{coleman2019}
Coleman G. A.~L.,  Leleu A.,  Alibert Y.,   Benz W.,  2019, \mn@doi [Astronomy
  and Astrophysics] {10.1051/0004-6361/201935922}, 631, A7

\bibitem[\protect\citeauthoryear{Collier~Cameron, Bruce, Miller, Triaud  \&
  Queloz}{Collier~Cameron et~al.}{2010}]{colliercameron2010}
Collier~Cameron A.,  Bruce V.~A.,  Miller G. R.~M.,  Triaud A. H. M.~J.,
  Queloz D.,  2010, \mn@doi [Monthly Notices of the Royal Astronomical Society]
  {10.1111/j.1365-2966.2009.16131.x}, 403, 151

\bibitem[\protect\citeauthoryear{Damasso et~al.,}{Damasso
  et~al.}{2020}]{damasso2020}
Damasso M.,  et~al., 2020, arXiv e-prints, 2007, arXiv:2007.06410

\bibitem[\protect\citeauthoryear{Dawson}{Dawson}{2014}]{dawson2014}
Dawson R.~I.,  2014, \mn@doi [The Astrophysical Journal]
  {10.1088/2041-8205/790/2/L31}, 790, L31

\bibitem[\protect\citeauthoryear{Dawson, Lee  \& Chiang}{Dawson
  et~al.}{2016}]{dawson2016}
Dawson R.~I.,  Lee E.~J.,   Chiang E.,  2016, \mn@doi [The Astrophysical
  Journal] {10.3847/0004-637X/822/1/54}, 822, 54

\bibitem[\protect\citeauthoryear{De~Rosa, Dawson  \& Nielsen}{De~Rosa
  et~al.}{2020}]{derosa2020}
De~Rosa R.~J.,  Dawson R.,   Nielsen E.~L.,  2020, arXiv e-prints, 2007,
  arXiv:2007.08549

\bibitem[\protect\citeauthoryear{Dressing \& Charbonneau}{Dressing \&
  Charbonneau}{2015}]{dressing2015}
Dressing C.~D.,  Charbonneau D.,  2015, \mn@doi [The Astrophysical Journal]
  {10.1088/0004-637X/807/1/45}, 807, 45

\bibitem[\protect\citeauthoryear{Ehrenreich et~al.,}{Ehrenreich
  et~al.}{2020}]{ehrenreich2020}
Ehrenreich D.,  et~al., 2020, \mn@doi [Nature] {10.1038/s41586-020-2107-1},
  580, 597

\bibitem[\protect\citeauthoryear{Fabrycky \& Tremaine}{Fabrycky \&
  Tremaine}{2007}]{fabrycky2007}
Fabrycky D.,  Tremaine S.,  2007, \mn@doi [The Astrophysical Journal]
  {10.1086/521702}, 669, 1298

\bibitem[\protect\citeauthoryear{Farr et~al.,}{Farr et~al.}{2018}]{farr2018}
Farr W.~M.,  et~al., 2018, \mn@doi [The Astrophysical Journal]
  {10.3847/2041-8213/aadfde}, 865, L20

\bibitem[\protect\citeauthoryear{{Foreman-Mackey}}{{Foreman-Mackey}}{2018}]{foreman-mackey2018}
{Foreman-Mackey} D.,  2018, \mn@doi [Research Notes of the American
  Astronomical Society] {10.3847/2515-5172/aaaf6c}, 2, 31

\bibitem[\protect\citeauthoryear{{Foreman-Mackey}}{{Foreman-Mackey}}{2019}]{foreman-mackey2019}
{Foreman-Mackey} D.,  2019, Astrophysics Source Code Library, p. ascl:1910.005

\bibitem[\protect\citeauthoryear{{Foreman-Mackey}, Hogg, Lang  \&
  Goodman}{{Foreman-Mackey} et~al.}{2013}]{foreman-mackey2013}
{Foreman-Mackey} D.,  Hogg D.~W.,  Lang D.,   Goodman J.,  2013, \mn@doi
  [Publications of the Astronomical Society of the Pacific] {10.1086/670067},
  125, 306

\bibitem[\protect\citeauthoryear{{Foreman-Mackey}, Agol, Ambikasaran  \&
  Angus}{{Foreman-Mackey} et~al.}{2017}]{foreman-mackey2017}
{Foreman-Mackey} D.,  Agol E.,  Ambikasaran S.,   Angus R.,  2017, \mn@doi [The
  Astronomical Journal] {10.3847/1538-3881/aa9332}, 154, 220

\bibitem[\protect\citeauthoryear{Fressin et~al.,}{Fressin
  et~al.}{2013}]{fressin2013}
Fressin F.,  et~al., 2013, \mn@doi [The Astrophysical Journal]
  {10.1088/0004-637X/766/2/81}, 766, 81

\bibitem[\protect\citeauthoryear{{Gaia Collaboration} et~al.,}{{Gaia
  Collaboration} et~al.}{2018}]{gaiacollaboration2018}
{Gaia Collaboration} et~al., 2018, \mn@doi [Astronomy and Astrophysics]
  {10.1051/0004-6361/201833051}, 616, A1

\bibitem[\protect\citeauthoryear{Gaidos, Mann, Kraus  \& Ireland}{Gaidos
  et~al.}{2016}]{gaidos2016}
Gaidos E.,  Mann A.~W.,  Kraus A.~L.,   Ireland M.,  2016, \mn@doi [Monthly
  Notices of the Royal Astronomical Society] {10.1093/mnras/stw097}, 457, 2877

\bibitem[\protect\citeauthoryear{Gandolfi et~al.,}{Gandolfi
  et~al.}{2018}]{gandolfi2018}
Gandolfi D.,  et~al., 2018, \mn@doi [Astronomy and Astrophysics]
  {10.1051/0004-6361/201834289}, 619, L10

\bibitem[\protect\citeauthoryear{Gelman, Carlin, Stern  \& Rubin}{Gelman
  et~al.}{2003}]{gelman2003}
Gelman A.,  Carlin J.~B.,  Stern H.~S.,   Rubin D.~B.,  2003, Bayesian {{Data
  Analysis}}, {{Second Edition}}.
{CRC Press}

\bibitem[\protect\citeauthoryear{Ghezzi, Cunha, Smith, {de Ara{\'u}jo}, Schuler
   \& {de la Reza}}{Ghezzi et~al.}{2010}]{ghezzi2010}
Ghezzi L.,  Cunha K.,  Smith V.~V.,  {de Ara{\'u}jo} F.~X.,  Schuler S.~C.,
  {de la Reza} R.,  2010, \mn@doi [The Astrophysical Journal]
  {10.1088/0004-637X/720/2/1290}, 720, 1290

\bibitem[\protect\citeauthoryear{Gillen, Hillenbrand, David, Aigrain, Rebull,
  Stauffer, Cody  \& Queloz}{Gillen et~al.}{2017}]{gillen2017}
Gillen E.,  Hillenbrand L.~A.,  David T.~J.,  Aigrain S.,  Rebull L.,  Stauffer
  J.,  Cody A.~M.,   Queloz D.,  2017, \mn@doi [The Astrophysical Journal]
  {10.3847/1538-4357/aa84b3}, 849, 11

\bibitem[\protect\citeauthoryear{Grunblatt, Howard  \& Haywood}{Grunblatt
  et~al.}{2015}]{grunblatt2015}
Grunblatt S.~K.,  Howard A.~W.,   Haywood R.~D.,  2015, \mn@doi [The
  Astrophysical Journal] {10.1088/0004-637X/808/2/127}, 808, 127

\bibitem[\protect\citeauthoryear{Hansen \& Murray}{Hansen \&
  Murray}{2012}]{hansen2012}
Hansen B. M.~S.,  Murray N.,  2012, \mn@doi [The Astrophysical Journal]
  {10.1088/0004-637X/751/2/158}, 751, 158

\bibitem[\protect\citeauthoryear{{Hardegree-Ullman}, Cushing, Muirhead  \&
  Christiansen}{{Hardegree-Ullman} et~al.}{2019}]{hardegree-ullman2019}
{Hardegree-Ullman} K.~K.,  Cushing M.~C.,  Muirhead P.~S.,   Christiansen
  J.~L.,  2019, \mn@doi [The Astronomical Journal] {10.3847/1538-3881/ab21d2},
  158, 75

\bibitem[\protect\citeauthoryear{Harvey}{Harvey}{1985}]{harvey1985a}
Harvey J.,  1985, in Rolfe E.,  Battrick B.,  eds,  {{ESA}} Special Publication
  Vol. 235, Future Missions in Solar, Heliospheric \& Space Plasma Physics.
  p.~199

\bibitem[\protect\citeauthoryear{Haywood et~al.,}{Haywood
  et~al.}{2014}]{haywood2014}
Haywood R.~D.,  et~al., 2014, \mn@doi [Monthly Notices of the Royal
  Astronomical Society] {10.1093/mnras/stu1320}, 443, 2517

\bibitem[\protect\citeauthoryear{Hirano, Narita, Shporer, Sato, Aoki  \&
  Tamura}{Hirano et~al.}{2011}]{hirano2011}
Hirano T.,  Narita N.,  Shporer A.,  Sato B.,  Aoki W.,   Tamura M.,  2011,
  \mn@doi [Publications of the Astronomical Society of Japan]
  {10.1093/pasj/63.sp2.S531}, 63, 531

\bibitem[\protect\citeauthoryear{Hirano et~al.,}{Hirano
  et~al.}{2020}]{hirano2020a}
Hirano T.,  et~al., 2020, \mn@doi [The Astrophysical Journal Letters]
  {10.3847/2041-8213/aba6eb}, 899, L13

\bibitem[\protect\citeauthoryear{Hoffman \& Gelman}{Hoffman \&
  Gelman}{2014}]{hoffman2014}
Hoffman M.~D.,  Gelman A.,  2014, Journal of Machine Learning Research, 15,
  1593

\bibitem[\protect\citeauthoryear{Howard et~al.,}{Howard
  et~al.}{2010}]{howard2010}
Howard A.~W.,  et~al., 2010, \mn@doi [Science] {10.1126/science.1194854}, 330,
  653

\bibitem[\protect\citeauthoryear{Huang et~al.,}{Huang
  et~al.}{2018a}]{huang2018a}
Huang C.~X.,  et~al., 2018a, arXiv e-prints, p. arXiv:1807.11129

\bibitem[\protect\citeauthoryear{Huang et~al.,}{Huang
  et~al.}{2018b}]{huang2018}
Huang C.~X.,  et~al., 2018b, \mn@doi [The Astrophysical Journal Letters]
  {10.3847/2041-8213/aaef91}, 868, L39

\bibitem[\protect\citeauthoryear{Hunter}{Hunter}{2007}]{hunter2007a}
Hunter J.~D.,  2007, \mn@doi [Computing in Science Engineering]
  {10.1109/MCSE.2007.55}, 9, 90

\bibitem[\protect\citeauthoryear{Johansen \& Lambrechts}{Johansen \&
  Lambrechts}{2017}]{johansen2017}
Johansen A.,  Lambrechts M.,  2017, \mn@doi [Annual Review of Earth and
  Planetary Sciences] {10.1146/annurev-earth-063016-020226}, 45, 359

\bibitem[\protect\citeauthoryear{Jones, Paul~Butler, Tinney, Marcy, Penny,
  McCarthy, Carter  \& Pourbaix}{Jones et~al.}{2002}]{jones2002}
Jones H. R.~A.,  Paul~Butler R.,  Tinney C.~G.,  Marcy G.~W.,  Penny A.~J.,
  McCarthy C.,  Carter B.~D.,   Pourbaix D.,  2002, \mn@doi [Monthly Notices of
  the Royal Astronomical Society] {10.1046/j.1365-8711.2002.05459.x}, 333, 871

\bibitem[\protect\citeauthoryear{Kipping}{Kipping}{2013}]{kipping2013a}
Kipping D.~M.,  2013, \mn@doi [Monthly Notices of the Royal Astronomical
  Society] {10.1093/mnrasl/slt075}, 434, L51

\bibitem[\protect\citeauthoryear{Kostov et~al.,}{Kostov
  et~al.}{2020}]{kostov2020}
Kostov V.~B.,  et~al., 2020, arXiv e-prints, 2004, arXiv:2004.07783

\bibitem[\protect\citeauthoryear{{Kunovac-Hodzic} \& Triaud}{{Kunovac-Hodzic}
  \& Triaud}{2019}]{kunovac-hodzic2019}
{Kunovac-Hodzic} V.,  Triaud A.,  2019, AAS - Extreme Solar Systems IV, 4,
  308.01

\bibitem[\protect\citeauthoryear{Kunovac~Hod{\v z}i{\'c}
  et~al.,}{Kunovac~Hod{\v z}i{\'c} et~al.}{2020}]{kunovachodzic2020}
Kunovac~Hod{\v z}i{\'c} V.,  et~al., 2020, arXiv e-prints, 2007,
  arXiv:2007.05514

\bibitem[\protect\citeauthoryear{Lambrechts, Morbidelli, Jacobson, Johansen,
  Bitsch, Izidoro  \& Raymond}{Lambrechts et~al.}{2019}]{lambrechts2019}
Lambrechts M.,  Morbidelli A.,  Jacobson S.~A.,  Johansen A.,  Bitsch B.,
  Izidoro A.,   Raymond S.~N.,  2019, \mn@doi [Astronomy and Astrophysics]
  {10.1051/0004-6361/201834229}, 627, A83

\bibitem[\protect\citeauthoryear{Lee, Chiang  \& Ormel}{Lee
  et~al.}{2014}]{lee2014}
Lee E.~J.,  Chiang E.,   Ormel C.~W.,  2014, \mn@doi [The Astrophysical
  Journal] {10.1088/0004-637X/797/2/95}, 797, 95

\bibitem[\protect\citeauthoryear{{Lightkurve Collaboration}
  et~al.,}{{Lightkurve Collaboration}
  et~al.}{2018}]{lightkurvecollaboration2018}
{Lightkurve Collaboration} et~al., 2018, Astrophysics Source Code Library, p.
  ascl:1812.013

\bibitem[\protect\citeauthoryear{{L{\'o}pez-Morales}
  et~al.,}{{L{\'o}pez-Morales} et~al.}{2014}]{lopez-morales2014}
{L{\'o}pez-Morales} M.,  et~al., 2014, \mn@doi [The Astrophysical Journal]
  {10.1088/2041-8205/792/2/L31}, 792, L31

\bibitem[\protect\citeauthoryear{Luger, Agol, {Foreman-Mackey}, Fleming,
  {Lustig-Yaeger}  \& Deitrick}{Luger et~al.}{2019}]{luger2019}
Luger R.,  Agol E.,  {Foreman-Mackey} D.,  Fleming D.~P.,  {Lustig-Yaeger} J.,
   Deitrick R.,  2019, \mn@doi [The Astronomical Journal]
  {10.3847/1538-3881/aae8e5}, 157, 64

\bibitem[\protect\citeauthoryear{Martin}{Martin}{2018}]{martin2018}
Martin D.~V.,  2018, \mn@doi [Handbook of Exoplanets]
  {10.1007/978-3-319-55333-7_156}, p.~156

\bibitem[\protect\citeauthoryear{Masuda \& Winn}{Masuda \&
  Winn}{2020}]{masuda2020}
Masuda K.,  Winn J.~N.,  2020, \mn@doi [The Astronomical Journal]
  {10.3847/1538-3881/ab65be}, 159, 81

\bibitem[\protect\citeauthoryear{Matsumura, Thommes, Chatterjee  \&
  Rasio}{Matsumura et~al.}{2010}]{matsumura2010}
Matsumura S.,  Thommes E.~W.,  Chatterjee S.,   Rasio F.~A.,  2010, \mn@doi
  [The Astrophysical Journal] {10.1088/0004-637X/714/1/194}, 714, 194

\bibitem[\protect\citeauthoryear{Maxted}{Maxted}{2016}]{maxted2016}
Maxted P. F.~L.,  2016, \mn@doi [Astronomy and Astrophysics]
  {10.1051/0004-6361/201628579}, 591, A111

\bibitem[\protect\citeauthoryear{Mayor et~al.,}{Mayor et~al.}{2011}]{mayor2011}
Mayor M.,  et~al., 2011, arXiv e-prints, 1109, arXiv:1109.2497

\bibitem[\protect\citeauthoryear{McLaughlin}{McLaughlin}{1924}]{mclaughlin1924}
McLaughlin D.~B.,  1924, \mn@doi [The Astrophysical Journal] {10.1086/142826},
  60, 22

\bibitem[\protect\citeauthoryear{Mohanty, Jankovic, Tan  \& Owen}{Mohanty
  et~al.}{2018}]{mohanty2018}
Mohanty S.,  Jankovic M.~R.,  Tan J.~C.,   Owen J.~E.,  2018, \mn@doi [The
  Astrophysical Journal] {10.3847/1538-4357/aabcd0}, 861, 144

\bibitem[\protect\citeauthoryear{Naoz, Farr, Lithwick, Rasio  \&
  Teyssandier}{Naoz et~al.}{2011}]{naoz2011a}
Naoz S.,  Farr W.~M.,  Lithwick Y.,  Rasio F.~A.,   Teyssandier J.,  2011,
  \mn@doi [Nature] {10.1038/nature10076}, 473, 187

\bibitem[\protect\citeauthoryear{Orosz et~al.,}{Orosz et~al.}{2019}]{orosz2019}
Orosz J.~A.,  et~al., 2019, \mn@doi [The Astronomical Journal]
  {10.3847/1538-3881/ab0ca0}, 157, 174

\bibitem[\protect\citeauthoryear{Paardekooper, Leinhardt, Th{\'e}bault  \&
  Baruteau}{Paardekooper et~al.}{2012}]{paardekooper2012}
Paardekooper S.-J.,  Leinhardt Z.~M.,  Th{\'e}bault P.,   Baruteau C.,  2012,
  \mn@doi [The Astrophysical Journal Letters] {10.1088/2041-8205/754/1/L16},
  754, L16

\bibitem[\protect\citeauthoryear{Palle et~al.,}{Palle et~al.}{2020}]{palle2020}
Palle E.,  et~al., 2020, \mn@doi [Astronomy and Astrophysics]
  {10.1051/0004-6361/202038583}, 643, A25

\bibitem[\protect\citeauthoryear{Pepe et~al.,}{Pepe et~al.}{2014}]{pepe2014}
Pepe F.,  et~al., 2014, \mn@doi [Astronomische Nachrichten]
  {10.1002/asna.201312004}, 335, 8

\bibitem[\protect\citeauthoryear{Pierens, McNally  \& Nelson}{Pierens
  et~al.}{2020}]{pierens2020}
Pierens A.,  McNally C.~P.,   Nelson R.~P.,  2020, \mn@doi [Monthly Notices of
  the Royal Astronomical Society] {10.1093/mnras/staa1550}, 496, 2849

\bibitem[\protect\citeauthoryear{Pollack, Hubickyj, Bodenheimer, Lissauer,
  Podolak  \& Greenzweig}{Pollack et~al.}{1996}]{pollack1996}
Pollack J.~B.,  Hubickyj O.,  Bodenheimer P.,  Lissauer J.~J.,  Podolak M.,
  Greenzweig Y.,  1996, \mn@doi [Icarus] {10.1006/icar.1996.0190}, 124, 62

\bibitem[\protect\citeauthoryear{Press \& Rybicki}{Press \&
  Rybicki}{1989}]{press1989}
Press W.~H.,  Rybicki G.~B.,  1989, \mn@doi [The Astrophysical Journal]
  {10.1086/167197}, 338, 277

\bibitem[\protect\citeauthoryear{Raymond, Boulet, Izidoro, Esteves  \&
  Bitsch}{Raymond et~al.}{2018}]{raymond2018}
Raymond S.~N.,  Boulet T.,  Izidoro A.,  Esteves L.,   Bitsch B.,  2018,
  \mn@doi [Monthly Notices of the Royal Astronomical Society]
  {10.1093/mnrasl/sly100}, 479, L81

\bibitem[\protect\citeauthoryear{Ricker et~al.,}{Ricker
  et~al.}{2015}]{ricker2015}
Ricker G.~R.,  et~al., 2015, \mn@doi [Journal of Astronomical Telescopes,
  Instruments, and Systems] {10.1117/1.JATIS.1.1.014003}, 1, 014003

\bibitem[\protect\citeauthoryear{Rossiter}{Rossiter}{1924}]{rossiter1924}
Rossiter R.~A.,  1924, \mn@doi [The Astrophysical Journal] {10.1086/142825},
  60, 15

\bibitem[\protect\citeauthoryear{Salvatier, Wiecki  \& Fonnesbeck}{Salvatier
  et~al.}{2016}]{salvatier2016}
Salvatier J.,  Wiecki T.~V.,   Fonnesbeck C.,  2016, \mn@doi [PeerJ Computer
  Science] {10.7717/peerj-cs.55}, 2, e55

\bibitem[\protect\citeauthoryear{Schlichting}{Schlichting}{2014}]{schlichting2014}
Schlichting H.~E.,  2014, \mn@doi [The Astrophysical Journal]
  {10.1088/2041-8205/795/1/L15}, 795, L15

\bibitem[\protect\citeauthoryear{Schlichting}{Schlichting}{2018}]{schlichting2018}
Schlichting H.~E.,  2018, \mn@doi [arXiv:1802.03090 [astro-ph]]
  {10.1007/978-3-319-30648-3\_141-1}, pp 1--20

\bibitem[\protect\citeauthoryear{Schoonenberg, Liu, Ormel  \&
  Dorn}{Schoonenberg et~al.}{2019}]{schoonenberg2019}
Schoonenberg D.,  Liu B.,  Ormel C.~W.,   Dorn C.,  2019, \mn@doi [Astronomy
  and Astrophysics] {10.1051/0004-6361/201935607}, 627, A149

\bibitem[\protect\citeauthoryear{Team et~al.,}{Team
  et~al.}{2016}]{thetheanodevelopmentteam2016}
Team T. T.~D.,  et~al., 2016, arXiv:1605.02688 [cs]

\bibitem[\protect\citeauthoryear{Triaud}{Triaud}{2018}]{triaud2018}
Triaud A. H. M.~J.,  2018, \mn@doi [Handbook of Exoplanets]
  {10.1007/978-3-319-55333-7_2}, p.~2

\bibitem[\protect\citeauthoryear{Valenti \& Fischer}{Valenti \&
  Fischer}{2005}]{valenti2005}
Valenti J.~A.,  Fischer D.~A.,  2005, \mn@doi [The Astrophysical Journal
  Supplement Series] {10.1086/430500}, 159, 141

\bibitem[\protect\citeauthoryear{Winn et~al.,}{Winn et~al.}{2010a}]{winn2010c}
Winn J.~N.,  et~al., 2010a, \mn@doi [The Astrophysical Journal]
  {10.1088/0004-637X/718/1/575}, 718, 575

\bibitem[\protect\citeauthoryear{Winn, Fabrycky, Albrecht  \& Johnson}{Winn
  et~al.}{2010b}]{winn2010a}
Winn J.~N.,  Fabrycky D.,  Albrecht S.,   Johnson J.~A.,  2010b, \mn@doi [The
  Astrophysical Journal] {10.1088/2041-8205/718/2/L145}, 718, L145

\bibitem[\protect\citeauthoryear{Wu}{Wu}{2019}]{wu2019}
Wu Y.,  2019, \mn@doi [The Astrophysical Journal] {10.3847/1538-4357/ab06f8},
  874, 91

\bibitem[\protect\citeauthoryear{Wu \& Lithwick}{Wu \& Lithwick}{2011}]{wu2011}
Wu Y.,  Lithwick Y.,  2011, \mn@doi [The Astrophysical Journal]
  {10.1088/0004-637X/735/2/109}, 735, 109

\bibitem[\protect\citeauthoryear{Xuan \& Wyatt}{Xuan \& Wyatt}{2020}]{xuan2020}
Xuan J.~W.,  Wyatt M.~C.,  2020, \mn@doi [Monthly Notices of the Royal
  Astronomical Society] {10.1093/mnras/staa2033}

\bibitem[\protect\citeauthoryear{Zurlo et~al.,}{Zurlo et~al.}{2018}]{zurlo2018}
Zurlo A.,  et~al., 2018, \mn@doi [Monthly Notices of the Royal Astronomical
  Society] {10.1093/mnras/sty1809}, 480, 35

\bibitem[\protect\citeauthoryear{{van der Walt}, Colbert  \& Varoquaux}{{van
  der Walt} et~al.}{2011}]{vanderwalt2011}
{van der Walt} S.,  Colbert S.~C.,   Varoquaux G.,  2011, \mn@doi [Computing in
  Science and Engineering] {10.1109/MCSE.2011.37}, 13, 22

\makeatother
\end{thebibliography}




\appendix


\section{\tess{} photometric analysis}
\label{appendix:tess}

The most recent published orbital parameters for \pimenc{} are based on an analysis of
\tess{} Sector 1 data \citep{huang2018,gandolfi2018}, with respectively 5 and 7 minute
uncertainties on the ephemerides at the time of our spectroscopic transits.
Moreover, the initial analyses did not take into account correlated noise in
their modelling, which may lead to underestimated errors on the reported transit
parameters and can impact our Rossiter-McLaughlin modelling. The light
curve from Sector 1 may also show evidence of rotational modulation from a spot on
the stellar surface, which may bias the measurement of the transit depth to
higher values and thus overestimate the Rossiter-McLaughlin amplitude.

\pimen{} is in a region of the sky that is being overlapped by several
\tess{} 
sectors while the spacecraft surveyed the Southern Hemisphere in Cycle 1, which
has since completed. \pimen{} has been
observed in Sectors 1 (25 July -- 22 August 2018); 4 (18 October  --
15 November 2018); 8 (2 -- 28 February 2019), 11 (22 April -- 21 May 2019), 12
(21 May -- 19 June 2019); and 13 (19 June -- 18 July 2019). 
In this section we outline our light curve
modelling including the five remaining \tess{} sectors that were not yet available in the discovery
papers, and take into account correlated noise using a Gaussian process model coupled
to a transit model.

\subsection{Flux extraction and pixel-level decorrelation}
\label{section:observations_photometry}

The \tess{} 2-minute cadence Target Pixel Files (TPF) for sectors 1, 4, 8, 11, 12,
and 13 
were downloaded using the \textsc{lightkurve} package
\citep{lightkurvecollaboration2018}. 
We
performed simple aperture photometry using the optimal aperture as determined by
the \tess{} Science Processing Operations Center (SPOC). Using this aperture, the
relative contribution from contaminant flux in our target aperture is of the
order \SI{0.3}{\ppt}, which has a negligible impact (${\sim}\SI{10}{\ppm}$) 
on the transit signals. 

The raw light curves have systematic effects from
the instrument and telescope motion, in particular at the beginning and ends of
each \tess{} orbit. In order to remove these effects we applied pixel-level decorrelation (PLD) on the
raw light curves using
the first order PLD basis and the top principal components from the second order
basis.  
In this work we opted for an implementation as detailed in the documentation for the
\exoplanet{}\footnote{\url{exoplanet.dfm.io}} package
\citep{foreman-mackey2019}, but a similar 
more flexible version can be found in the \textsc{lightkurve}\footnote{\url{http://docs.lightkurve.org}}
package as well. 

\subsection{Transit light curve modelling of \tess{} sectors 1, 4, 8, 11, 12, 13}
\label{section:light_curve_modelling}

We used the \exoplanet{} software package \citep{foreman-mackey2019} 
to model the \tess{} transit
signals, including a correlated noise model using Gaussian process regression
with \celerite{} \citep{foreman-mackey2017}. We assumed the star follows a quadratic limb darkening law and 
held the limb darkening parameters
$c_1 = 0.28$, $c_2 = 0.27$ fixed for the \tess{} band, following
\citep{huang2018}.
We fit for the orbital period,
$P$; transit midpoint, $T_0$; impact parameter, $b$; and planet radius,
$\ln{r_\mathrm{p}}$, while also sampling the stellar radius and mass with normal
priors 
$R_\star =
\SI{1.17 \pm 0.02}{\Rsun}$, $M_\star =
\SI{1.07 \pm 0.04}{\Msun}$ from the analysis in \citep{damasso2020}.

For our Gaussian process model we used a kernel with covariance function given by
\begin{align*}
    k(\tau) &= \sigma^2 \left[ (1 - 1/\epsilon) \exp{\left( -\frac{1 - \epsilon}{\rho}\sqrt{3} \tau \right )} (1 - 1/\epsilon) \exp{\left( -\frac{1 + \epsilon}{\rho}\sqrt{3} \tau \right)}\right] \\,
\end{align*}
where $\sigma$ describes the amplitude of the signal, and $\rho$ is a characteristic timescale. In the limit $\epsilon \to 0$ the above function becomes the Mat\'ern-$3/2$ function,
which can flexibly fit instrument 
systematics related to the \tess{} pointing as well as some astrophysical
variability, although the light curves seem to be dominated by the former. We use the default setting in \celerite{} to approximate the Mat\'ern-$3/2$ function ($\epsilon = 0.01$), and  fit for the logarithm of the amplitude, $\ln{\sigma}$, and logarithm
of the timescale, $\ln{\rho}$.
Finally, we also fit for the logarithm of a white noise term for our photometric uncertainties, $\ln{s^2}$,
as well as a photometric offset, $\Delta f$. Since each \tess{} sector has different noise properties, we fit each of the above nuisance parameters separately for each sector, while the transit parameters are shared between the full dataset. The full set of priors used in the analysis is shown in Table~\ref{table:transit_priors}.

We found that the measurement of a consistent transit depth between
\tess{} sectors is very sensitive to the
timescale of the baseline model being fit. Even after the PLD correction, some
data show somewhat disjoint ~2.5 day segments from 
``momentum dumps'' as thrusters are fired to reorient the \tess{} spacecraft,
pointing drifts at the start and/or end of each \tess{} orbit, and other
show timescale variation that is most likely related to the spacecraft pointing. In an
otherwise smoothly varying light curve, these ``rapid'' changes drive the
timescale of the Gaussian process kernel lower values to be able to fit both
high and low frequency variation, which will affect the transit depth. In order
to avoid biasing the measurement of the transit depth, we remove some
out-of-transit segments of
data associated with rapid changes in the light curve, which range
from a few hours to a few days. In addition, one transit from Sector 11 is
removed due to a momentum dump during transit that is not sufficiently corrected
from the PLD.

\begin{figure*}
    \centering
    \includegraphics{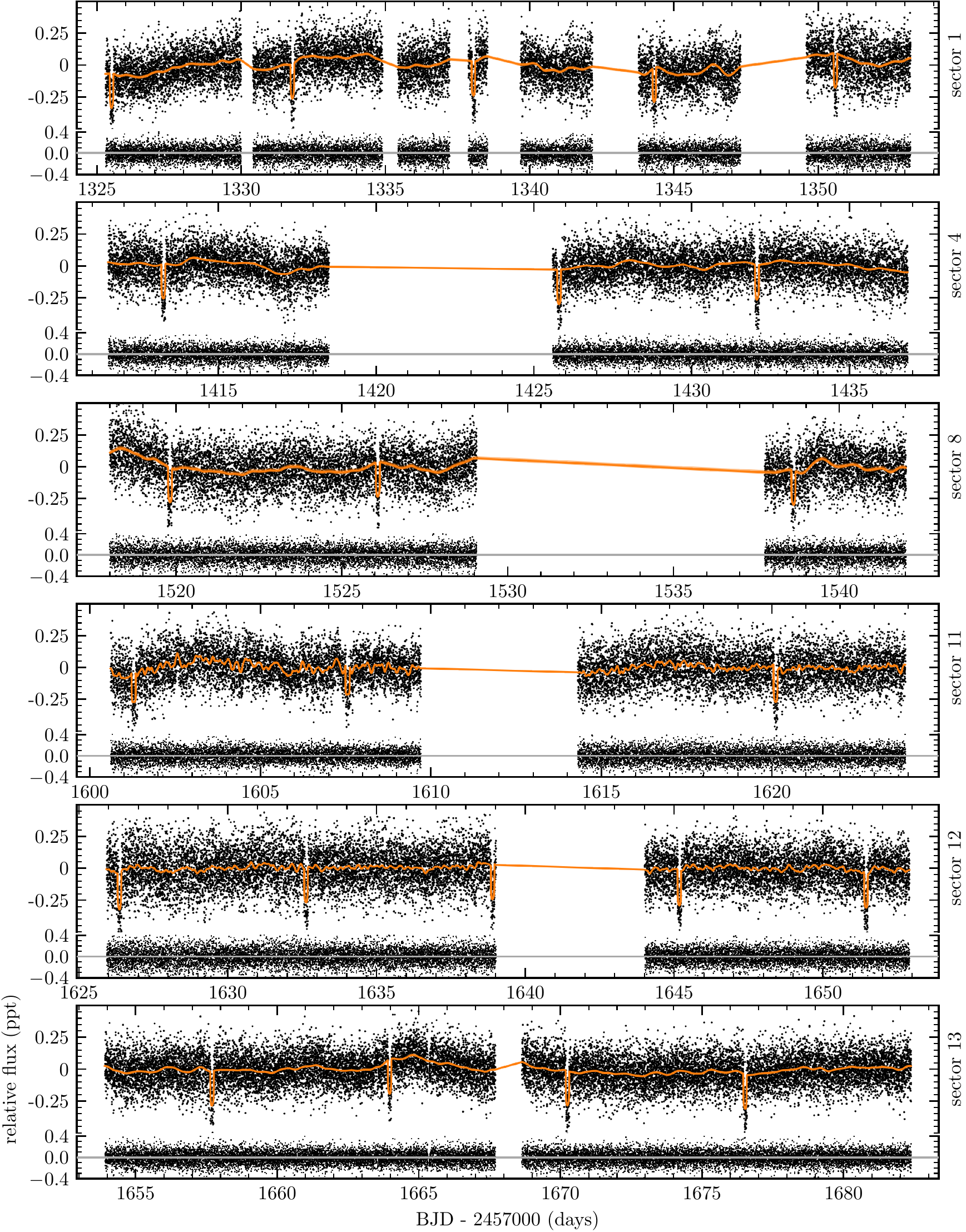}
    \caption{\tess data (\emph{black points}) from sectors 1, 4, 8, 11, 12, and
        13
        fitted with the Gaussian process model and transit model
        (\emph{orange line}) described in
         Section~\ref{appendix:tess}.
         In
         the lower panel of each sector we show the residuals from the best fit.}
    \label{fig:tess_fit}
\end{figure*}

We first fit the data using a least squares algorithm to find the maximum
likelihood, and performed a single
$5\sigma$-clip on the residuals from the best-fitting model to remove outliers.
We then sampled the free parameters of our model with Markov Chain Monte Carlo
using the No-U-Turn (NUTS) sampler
\citep{hoffman2014} as
implemented in \pymcthree{} \citep{salvatier2016}  
and {\exoplanet{}}. We launched four independent chains, discarded the first
\num{2000} tuning steps, and finally sampled \num{1000} additional 
steps, resulting in ${>}\num{1000}$ effective samples per parameter. 
All parameters reached the recommended $\hat{R} < 1.1$ (typically $\num{<1.001}$) convergence criterion 
\citep{gelman2003}. 
We performed the MCMC sampling individually for each \tess{} sector, and also using
the full combined dataset. The baseline-corrected, phase-folded transit light curve is shown in Fig.~\ref{fig:pimen_transit}, and the full transit and Gaussian process fit to the data is shown in Fig.~\ref{fig:tess_fit} with residual 
rms scatter of \SI{123}{\ppm}.

We also searched for transit timing variations (TTVs) by fitting individual transit times, but found no significant signal. Finally, we also searched the combined \tess{} residual light curve for solar-like oscillations, but
were unable to uncover evidence of detectable modes. However, as we report in
Section~\ref{section:espresso}, the high S/N \espresso{} data do show clearly detectable oscillations in Doppler velocity.

\begin{table}
\renewcommand{\arraystretch}{1.2}
    \centering
    \caption{Priors on the parameters for the photometric transit modelling. $N$ refers to individual sectors. $^{\textblue{a}}$Impact parameter prior (Beta distribution) from \citet{kipping2013a}.}
    \begin{tabular*}{\columnwidth}{@{\extracolsep{\fill}}
            lr}
        \toprule
        \toprule
        {Parameter} & {Prior} \\
        \midrule
        $M_\star$ (\si{\Msun}) & $\mathcal{N}(1.07, 0.04)$ \\
        $R_\star$ (\si{\Rsun}) & $\mathcal{N}(1.17, 0.02)$ \\
        $\ln{r_\mathrm{p}}$ (\si{\Rsun}) & $\mathcal{N}(-4.05, 1)$ \\
        $\ln{P}$ (days) & $\mathcal{N}(1.84, 1)$ \\
        $T_0$ (BJD-\num{2450000}) & $\mathcal{N}(8425.7892, 1)$ \\
        $b$ & $\mathcal{B}^{\textblue{a}}$ \\
        $\ln{\sigma_N}$ & $\mathcal{N}(\ln{\text{var}\,y_N}, 10)$ \\
        $\ln{\rho_N}$ & $\mathcal{N}(0, 10)$ \\
        $\ln{s^2_N}$ & $\mathcal{N}(\ln{\text{var}\,y_N}, 10)$ \\
        $\Delta f_N$ & $\mathcal{N}(1, 0.1)$ \\
        \bottomrule
    \end{tabular*}
    \label{table:transit_priors}
\end{table}

\section{\espresso{} observables}

\begin{figure}
    \centering
    \includegraphics[width=\linewidth]{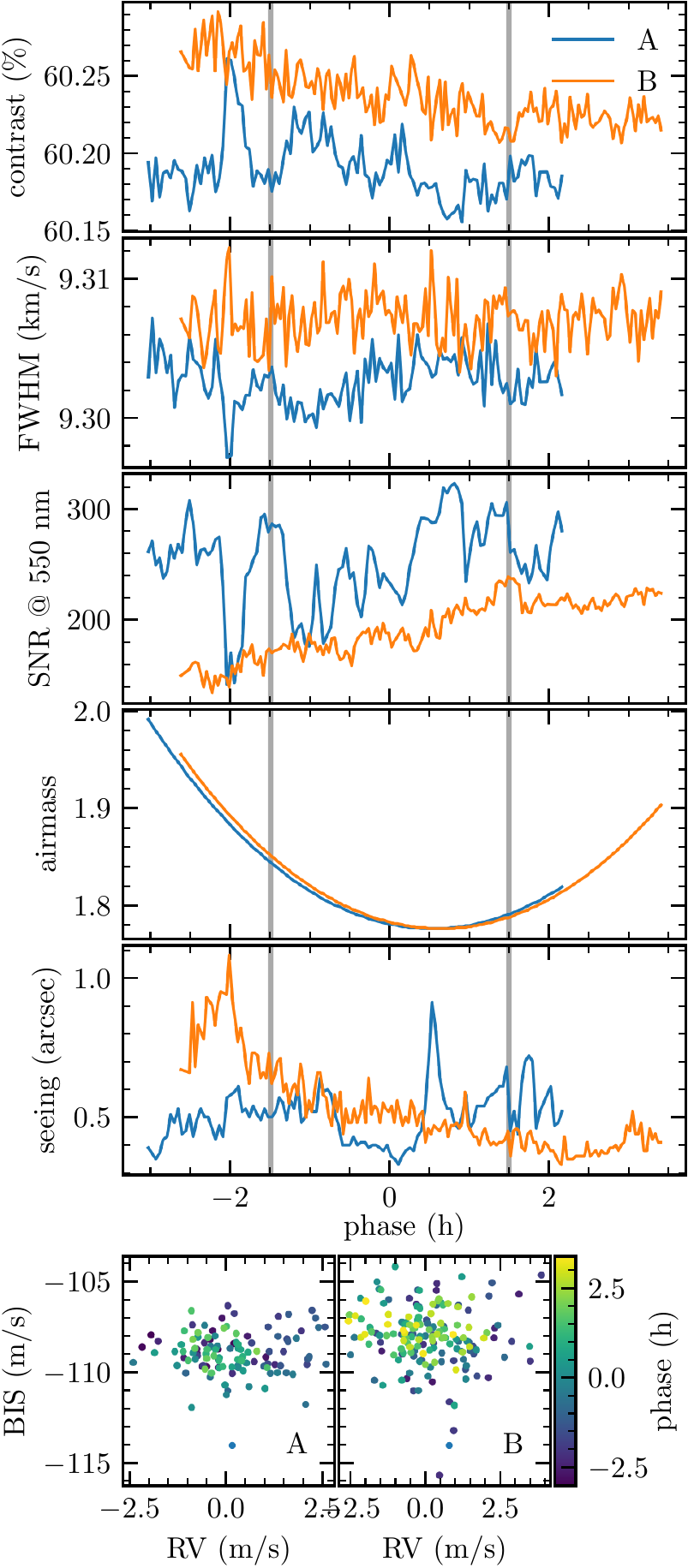}
    \caption{Observing conditions and CCF observables for \espresso{} run A (\emph{blue}) and run B (\emph{orange}). From top to bottom: CCF contrast, CCF FWHM, SNR at \SI{550}{\nm}, airmass, seeing, and bisector inverse slope (BIS). The grey vertical lines denote the transit ingress and egress.}
    \label{fig:observables}
\end{figure}

\section{Residual CCF profiles}
\label{appendix:ccfpo}

\begin{figure}
    \centering
    \includegraphics[width=0.9\linewidth]{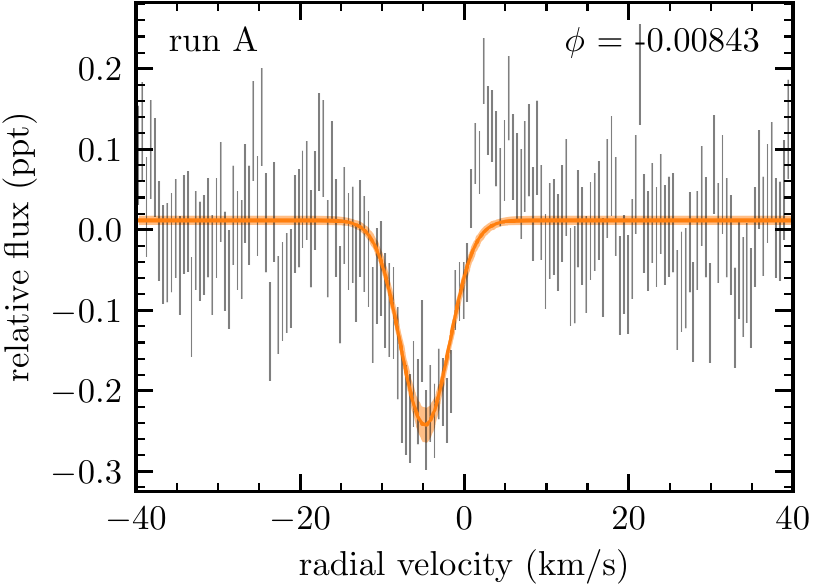}
    \includegraphics[width=0.9\linewidth]{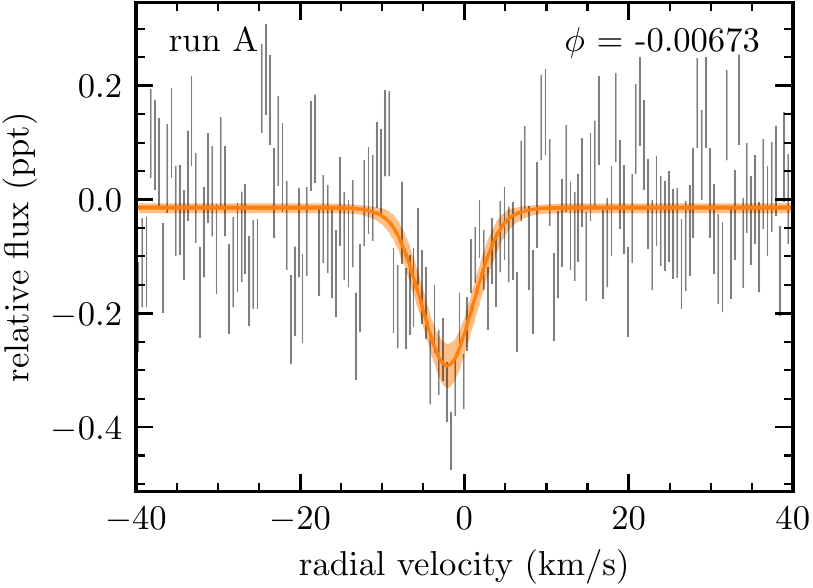}
    \includegraphics[width=0.9\linewidth]{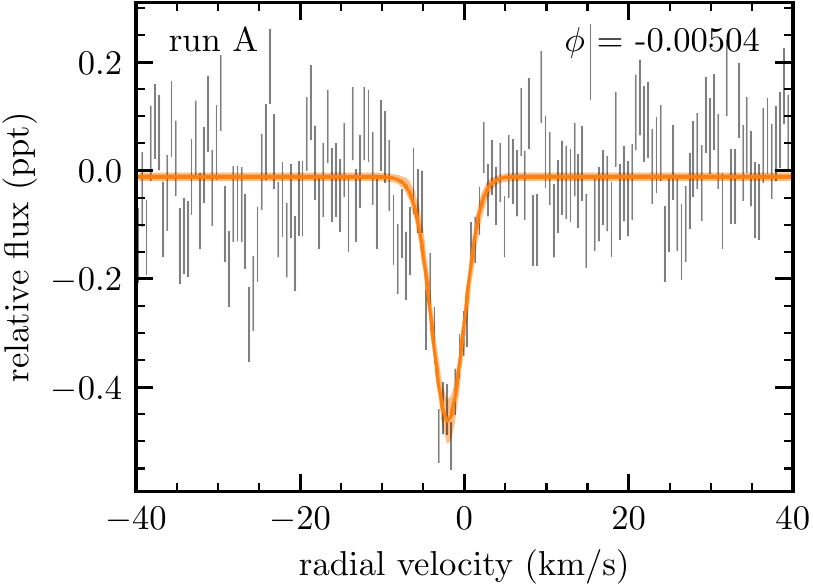}
    \includegraphics[width=0.9\linewidth]{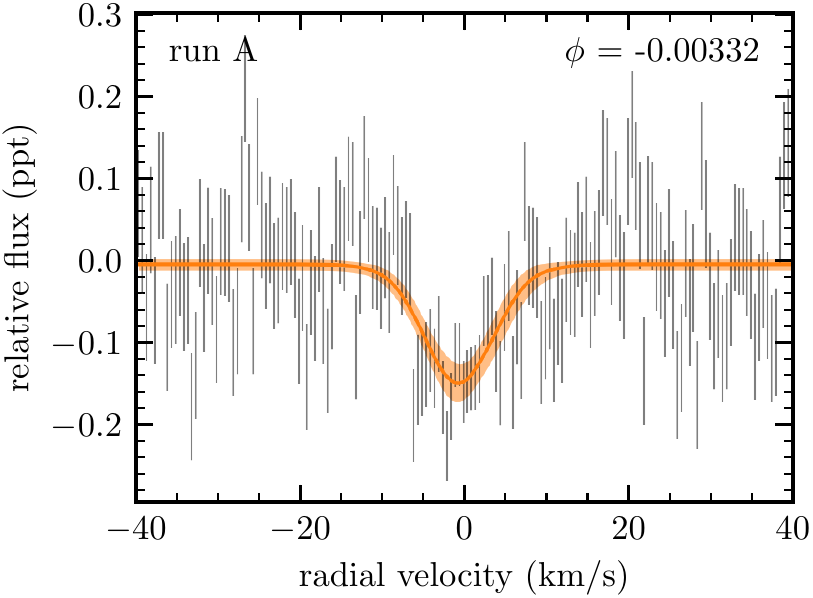}
    \caption{Gaussian fits to the 15 min binned residual line profiles from run A. The orange shading denotes the $1\sigma$ uncertainty of the model.}
    \label{fig:ccfpo1}
\end{figure}
\begin{figure}
    \centering
    \includegraphics[width=0.9\linewidth]{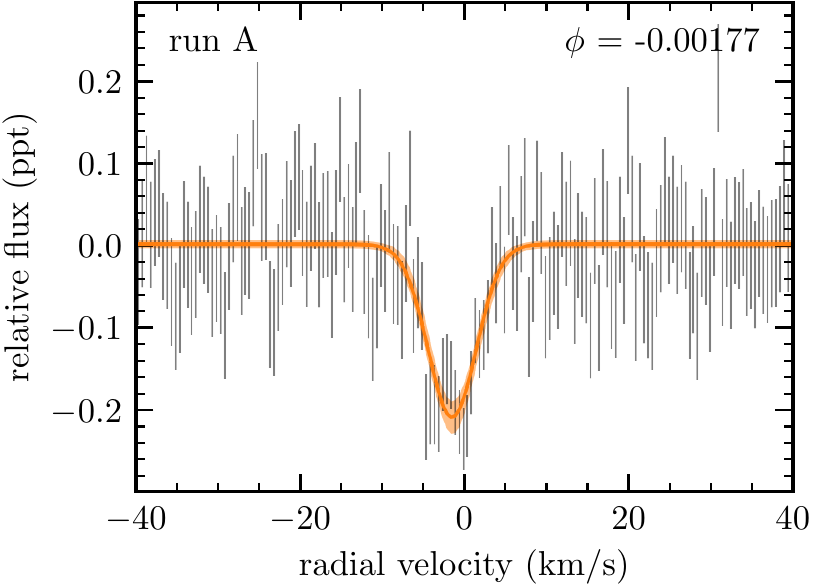}
    \includegraphics[width=0.9\linewidth]{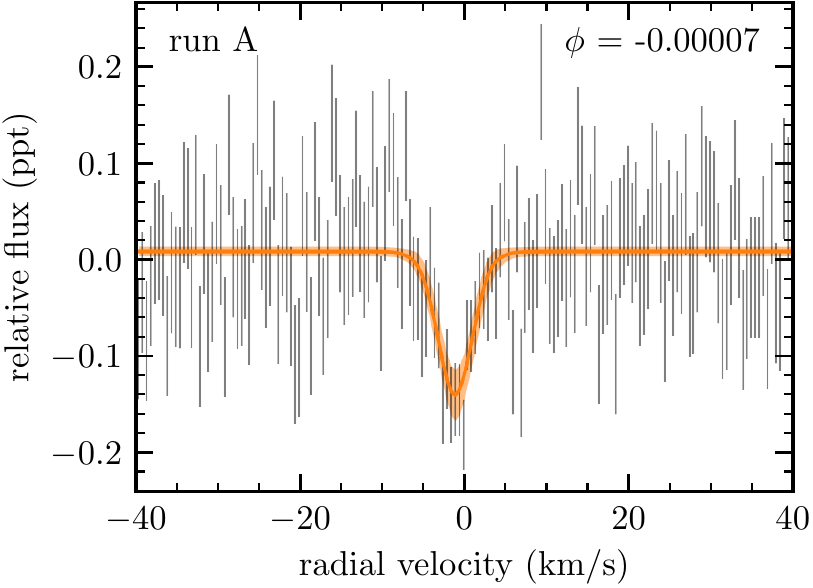}
    \includegraphics[width=0.9\linewidth]{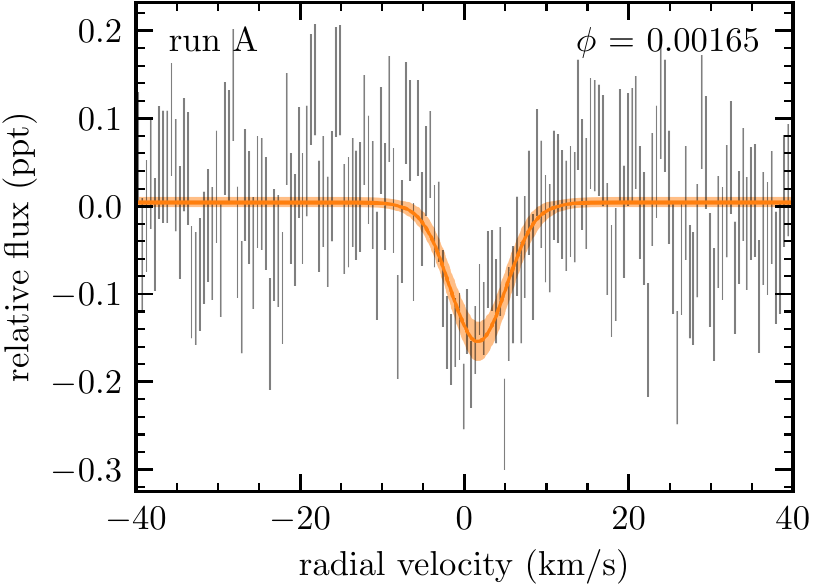}
    \includegraphics[width=0.9\linewidth]{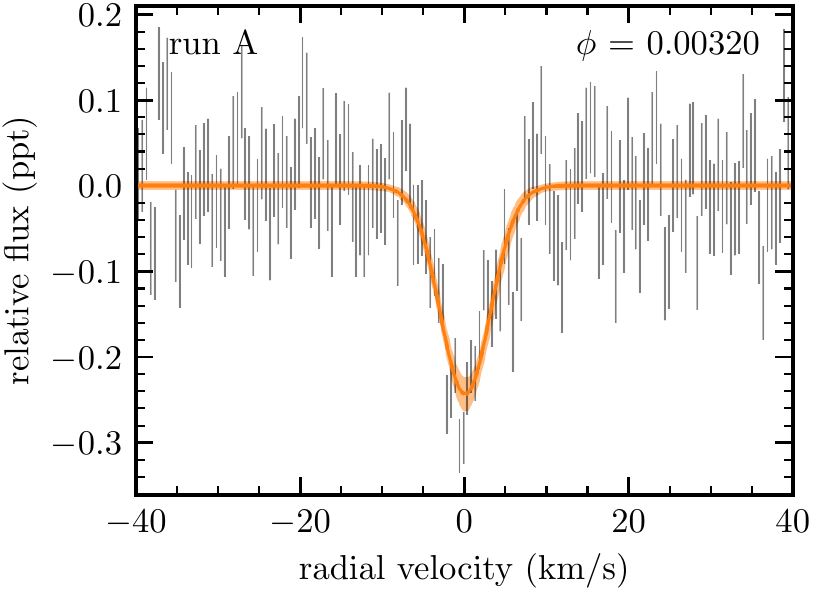}
    \caption{Gaussian fits to the 15 min binned residual line profiles from run A. The orange shading denotes the $1\sigma$ uncertainty of the model.}
    \label{fig:ccfpo2}
\end{figure}

\begin{figure}
    \centering
    \includegraphics[width=0.9\linewidth]{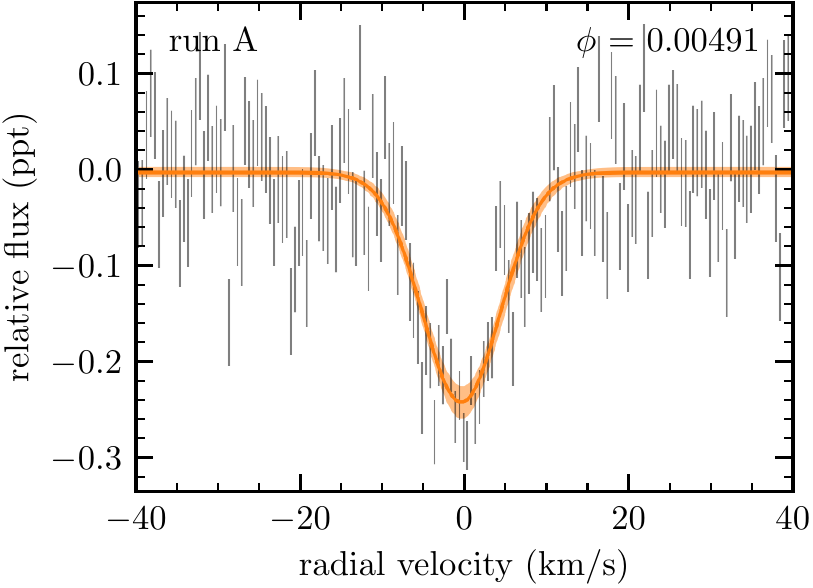}
    \includegraphics[width=0.9\linewidth]{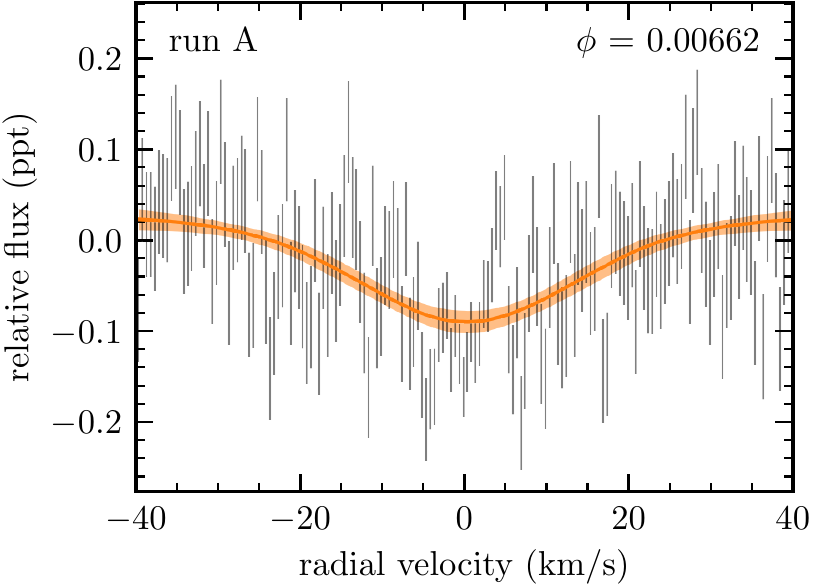}
    \includegraphics[width=0.9\linewidth]{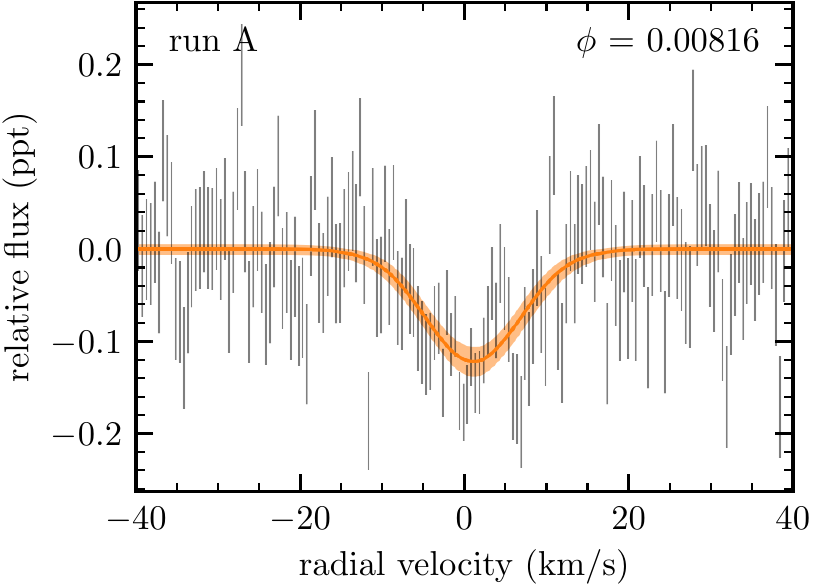}
    \includegraphics[width=0.9\linewidth]{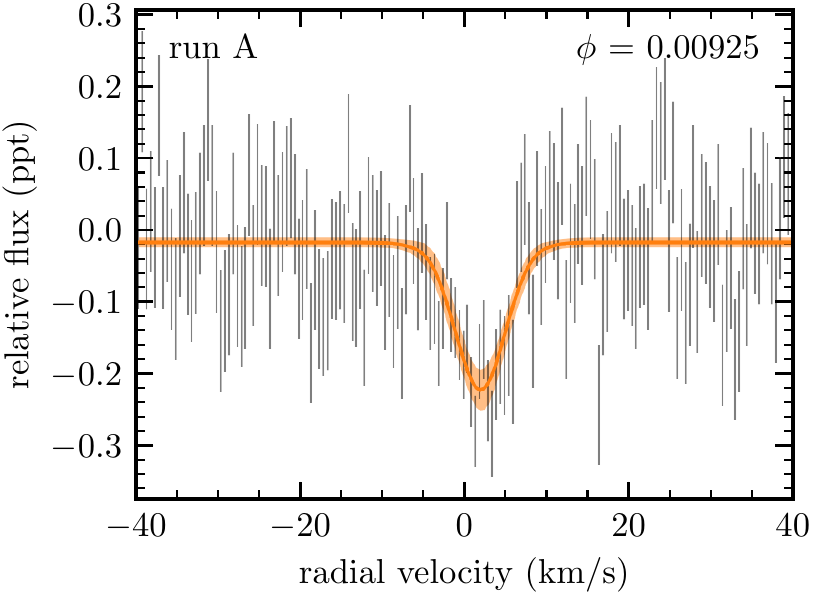}
    \caption{Gaussian fits to the 15 min binned residual line profiles from run A. The orange shading denotes the $1\sigma$ uncertainty of the model.}
    \label{fig:ccfpo3}
\end{figure}

\begin{figure}
    \centering
    \includegraphics[width=0.9\linewidth]{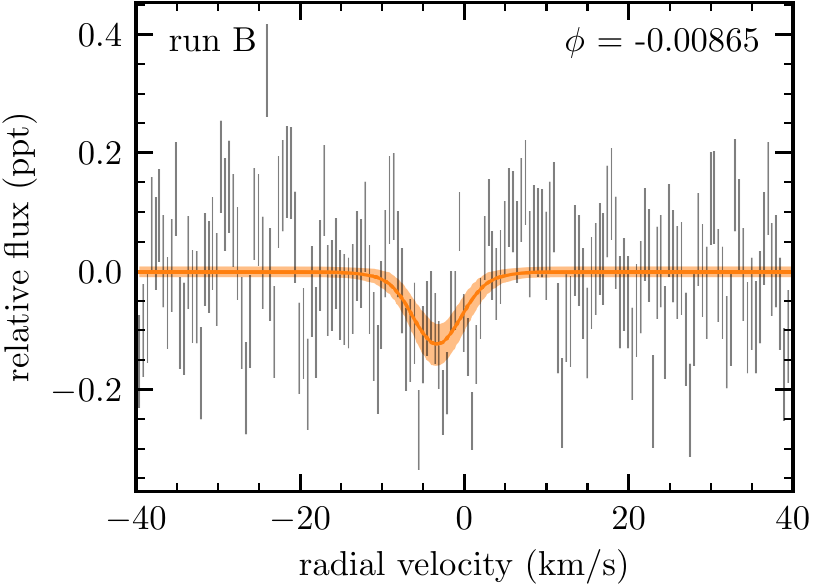}
    \includegraphics[width=0.9\linewidth]{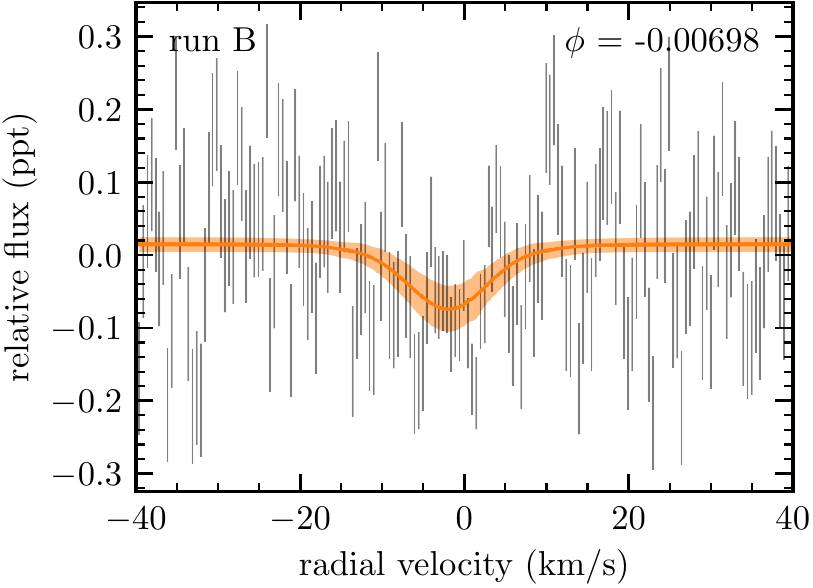}
    \includegraphics[width=0.9\linewidth]{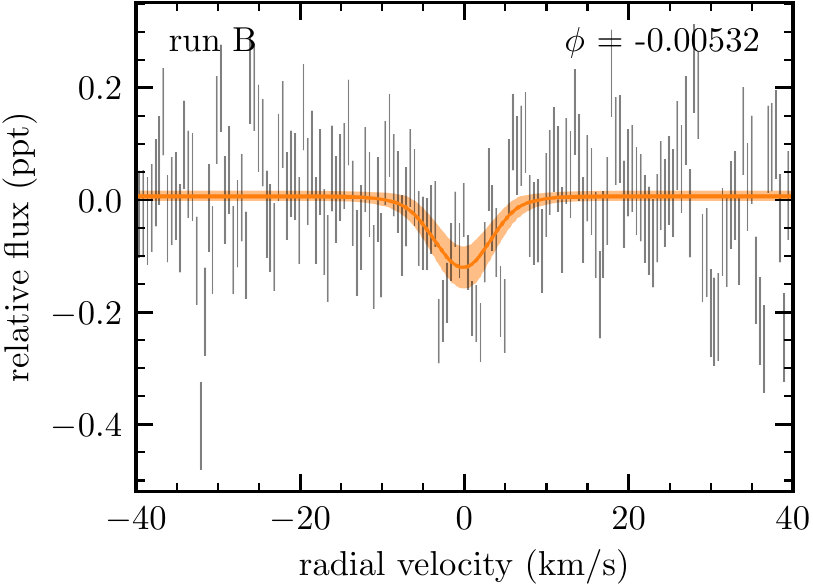}
    \includegraphics[width=0.9\linewidth]{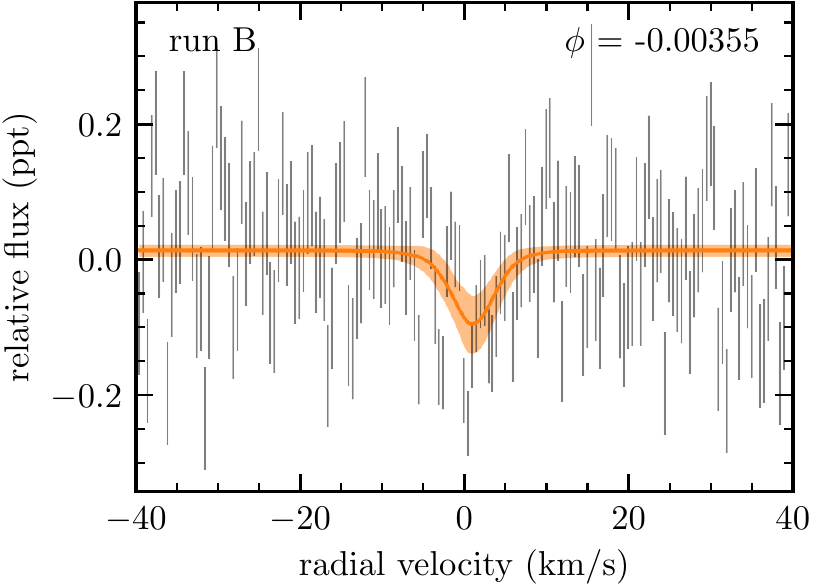}
    \caption{Gaussian fits to the 15 min binned residual line profiles from run B. The orange shading denotes the $1\sigma$ uncertainty of the model.}
    \label{fig:ccfpo4}
\end{figure}
\begin{figure}
    \centering
    \includegraphics[width=0.9\linewidth]{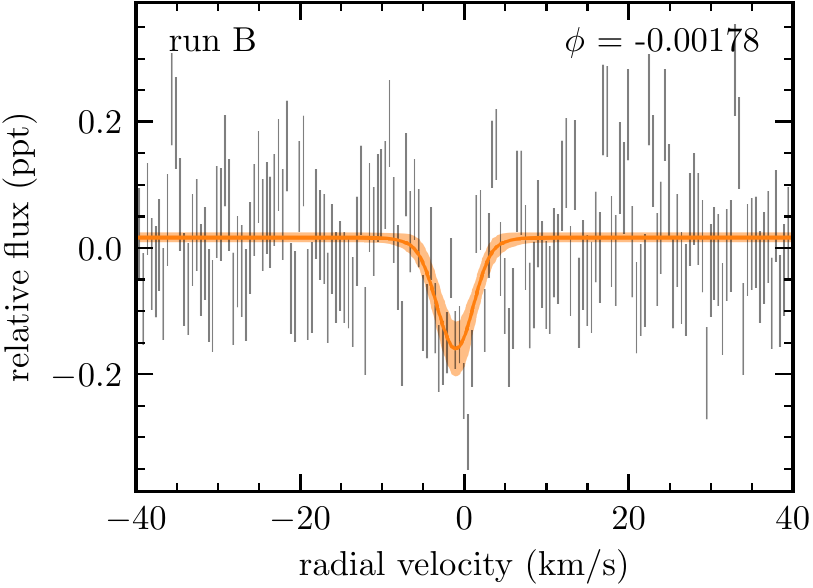}
    \includegraphics[width=0.9\linewidth]{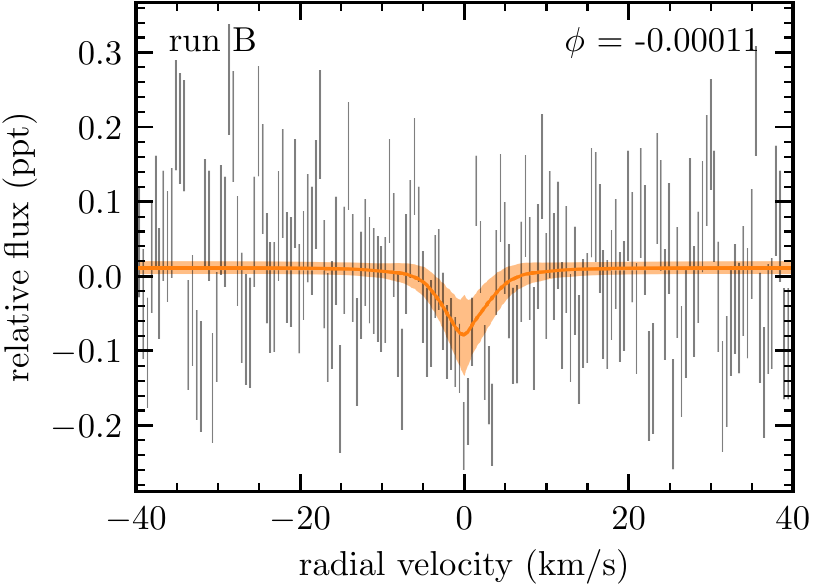}
    \includegraphics[width=0.9\linewidth]{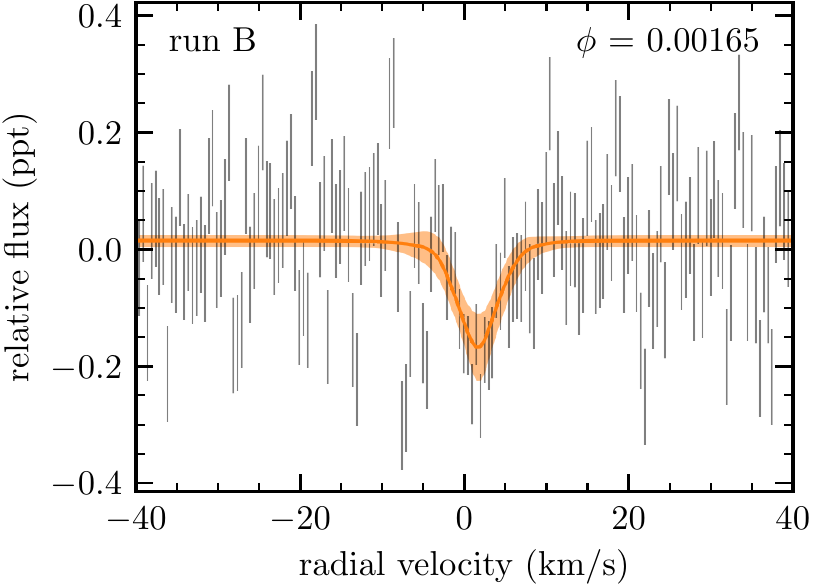}
    \includegraphics[width=0.9\linewidth]{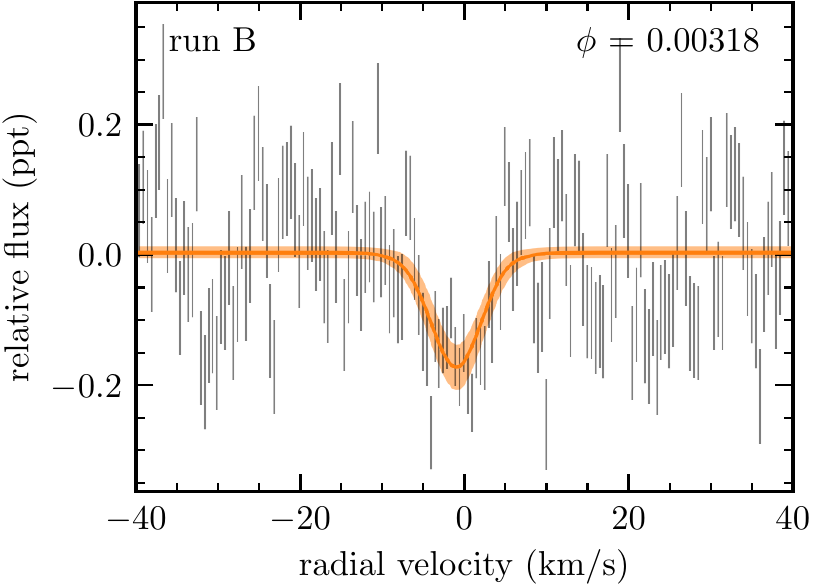}
    \caption{Gaussian fits to the 15 min binned residual line profiles from run B. The orange shading denotes the $1\sigma$ uncertainty of the model.}
    \label{fig:ccfpo5}
\end{figure}

\begin{figure}
    \centering
    \includegraphics[width=0.9\linewidth]{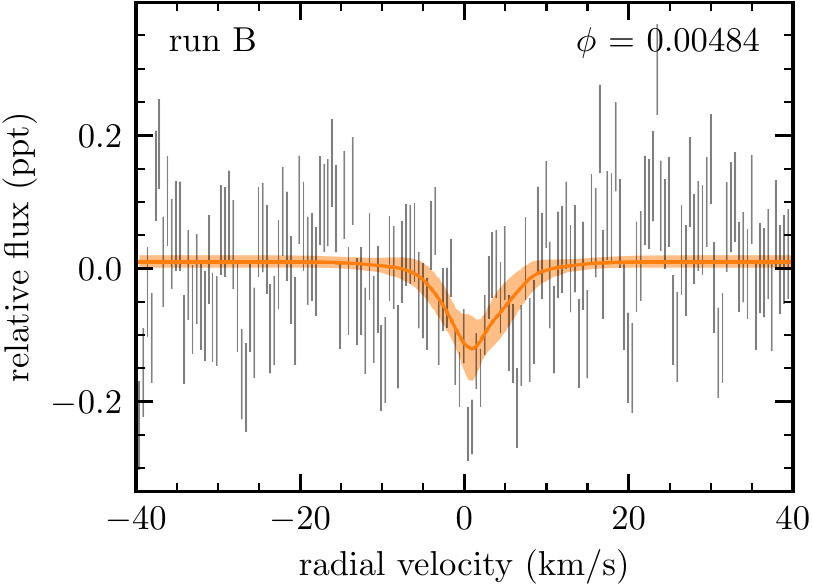}
    \includegraphics[width=0.9\linewidth]{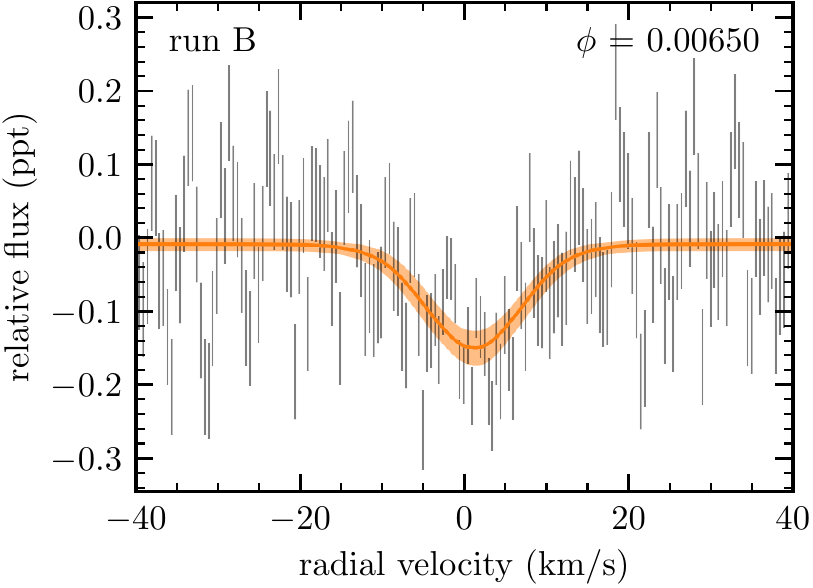}
    \includegraphics[width=0.9\linewidth]{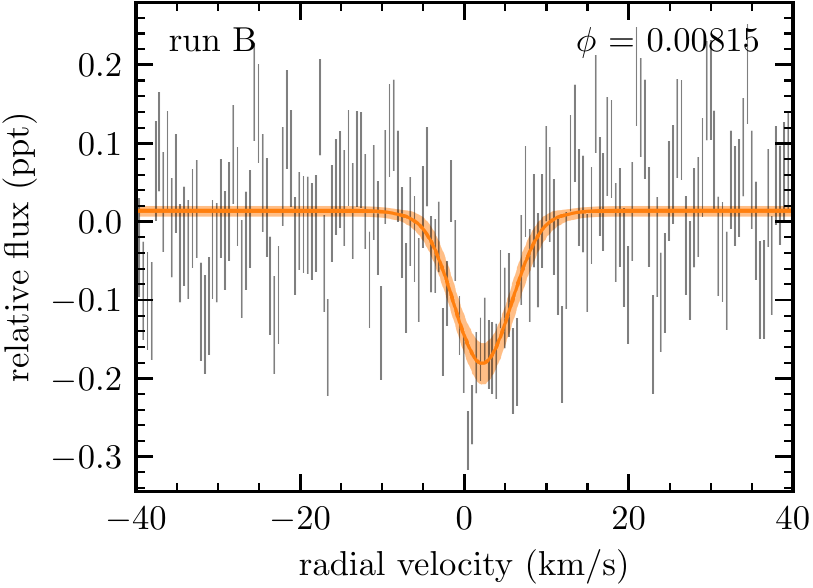}
    \includegraphics[width=0.9\linewidth]{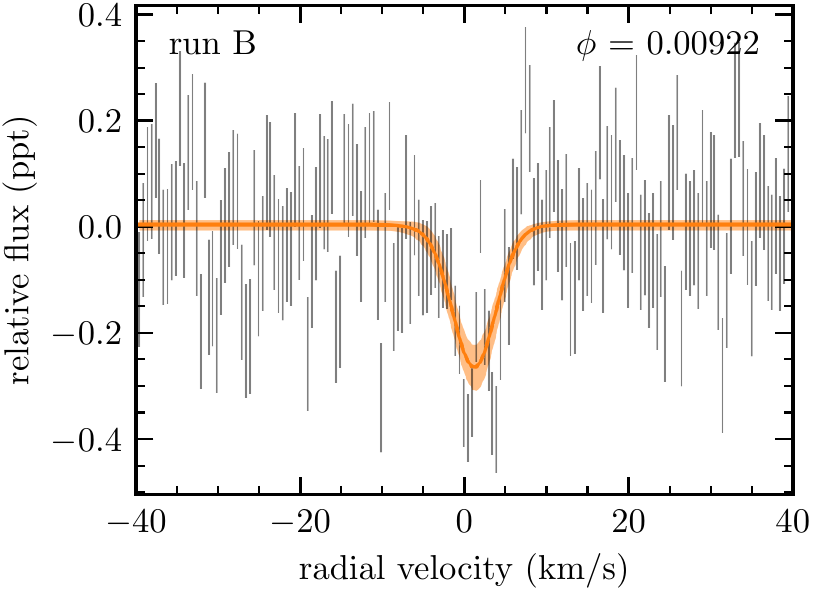}
    \caption{Gaussian fits to the 15 min binned residual line profiles from run B. The orange shading denotes the $1\sigma$ uncertainty of the model.}
    \label{fig:ccfpo6}
\end{figure}

\begin{figure}
    \centering
    \includegraphics{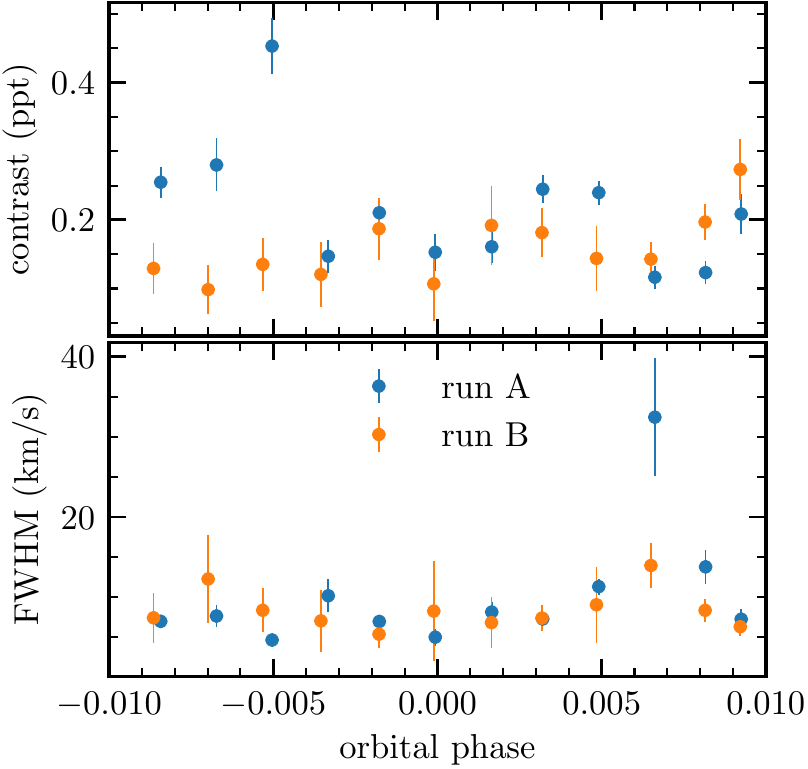}
    \caption{Comparison of the fitted local depth and width from the residual line profiles for run A and B.}
    \label{fig:local_fwhm_contrast}
\end{figure}

\section{Gaussian process hyperparameters}

\begin{table*}
\renewcommand{\arraystretch}{1.2}
    \centering
    \caption{Table of Gaussian process hyperparameters from the photometric and radial velocity analysis.}
    \begin{tabular*}{\linewidth}{@{\extracolsep{\fill}}
            llcc}
        \toprule
        \toprule
        {\textbf{Parameter}} & {\textbf{Description}} & {\textbf{Value}} & {\textbf{Source}}  \\
        \midrule
        \multicolumn{4}{@{}c@{}}{\emph{Photometric analysis}} \\
        \midrule
        \multicolumn{4}{@{}l@{}}{\emph{Sector 1}} \\
        $\log{\sigma}$ & Gaussian process amplitude & $-0.44^{+0.18}_{-0.17}$ & \tess{} \\
        $\log{\rho}$ & Gaussian process timescale & $-9.93^{+0.16}_{-0.13}$ & \tess{} \\
        $\log{s^2}$ & Flux variance & $-18.037^{+0.012}_{-0.012}$ & \tess{} \\
        $\Delta f$ & Mean flux & $1.0000054^{+0.0000111}_{-0.0000123}$ & \tess{} \\[5pt]
        \multicolumn{4}{@{}l@{}}{\emph{Sector 4}} \\
        $\log{\sigma}$ & Gaussian process amplitude & $-0.58^{+0.25}_{-0.26}$ & \tess{} \\
        $\log{\rho}$ & Gaussian process timescale & $-10.47^{+0.16}_{-0.14}$ & \tess{} \\
        $\log{s^2}$ & Flux variance & $-18.169^{+0.013}_{-0.012}$ & \tess{} \\
        $\Delta f$ & Mean flux & $0.9999993^{+0.0000078}_{-0.0000078}$ & \tess{} \\ [5pt]
        \multicolumn{4}{@{}l@{}}{\emph{Sector 8}} \\
        $\log{\sigma}$ & Gaussian process amplitude & $-0.47^{+0.26}_{-0.23}$ & \tess{} \\
        $\log{\rho}$ & Gaussian process timescale & $-9.95^{+0.18}_{-0.16}$ & \tess{} \\
        $\log{s^2}$ & Flux variance & $-17.903^{+0.013}_{-0.014}$ & \tess{} \\
        $\Delta f$ & Mean flux & $1.0000088^{+0.0000134}_{-0.0000139}$ & \tess{} \\[5pt]
        \multicolumn{4}{@{}l@{}}{\emph{Sector 11}} \\
        $\log{\sigma}$ & Gaussian process amplitude & $-2.63^{+0.15}_{-0.14}$ & \tess{} \\
        $\log{\rho}$ & Gaussian process timescale & $-10.26^{+0.06}_{-0.06}$ & \tess{} \\
        $\log{s^2}$ & Flux variance & $-18.083^{+0.012}_{-0.013}$ & \tess{} \\
        $\Delta f$ & Mean flux & $1.0000049^{+0.0000034}_{-0.0000033}$ & \tess{} \\[5pt]
        \multicolumn{4}{@{}l@{}}{\emph{Sector 12}} \\
        $\log{\sigma}$ & Gaussian process amplitude & $-2.52^{+0.28}_{-0.25}$ & \tess{} \\
        $\log{\rho}$ & Gaussian process timescale & $-10.71^{+0.08}_{-0.08}$ & \tess{} \\
        $\log{s^2}$ & Flux variance & $-17.873^{+0.012}_{-0.011}$ & \tess{} \\
        $\Delta f$ & Mean flux & $1.0000128^{+0.0000023}_{-0.0000023}$ & \tess{} \\[5pt]
        \multicolumn{4}{@{}l@{}}{\emph{Sector 13}} \\
        $\log{\sigma}$ & Gaussian process amplitude & $-0.64^{+0.23}_{-0.24}$ & \tess{} \\
        $\log{\rho}$ & Gaussian process timescale & $-10.41^{+0.12}_{-0.11}$ & \tess{} \\
        $\log{s^2}$ & Flux variance & $-18.211^{+0.011}_{-0.010}$ & \tess{} \\
        $\Delta f$ & Mean flux & $1.0000019^{+0.0000062}_{-0.0000060}$ & \tess{} \\[5pt]

        \multicolumn{4}{@{}c@{}}{\emph{Radial velocity analysis}} \\
        \midrule
        %
        $\log{S_\text{osc}}$ (\si{\kilo\metre\squared\per\second\squared}) & Oscillation power & $-22.58^{+0.21}_{-0.27}$ & \espresso{} RVs\\
        $\log{Q_\mathrm{osc}}$ & Oscillation damping & $1.58^{+0.23}_{-0.21}$ & \espresso{} RVs\\
        $\log{S_\text{bkg}}$ (\si{\kilo\metre\squared\per\second\squared}) & Background activity power & $-18.48^{+0.90}_{-0.68}$ & \espresso{} RVs\\
        $\log{\omega_\text{bkg}}$ (\si{\per\day}) & Background activity timescale & $4.07^{+0.32}_{-0.34}$ & \espresso{} RVs\\
        $\log{\sigma_\text{A}}$ (\si{\kilo\metre\per\second}) & White noise term for run A & $-10.31^{+1.25}_{-1.16}$ & \espresso{} RVs \\
        $\log{\sigma_\text{B}}$ (\si{\kilo\metre\per\second}) & White noise term for run B & $-8.29^{+0.67}_{-2.35}$ & \espresso{} RVs \\
        \bottomrule
    \end{tabular*}
    \label{table:results_hyperparameters}
\end{table*}

\bsp	
\label{lastpage}
\end{document}